\documentclass{ws-ijmpd}

\begin{document}

\markboth{Williams, Turyshev, Boggs}
{LLR Tests of the Equivalence Principle with the Earth and Moon}

\catchline{}{}{}{}{}
%
%%%%%%%%%%%%%%%%%%%%%%%%%%%%%%%%%%%%%%%%%%%%%%%%%%%%%%%%%%%%%%%%%%%%
%Lunar Laser Ranging Tests of the Equivalence Principle \\ with the Earth and Moon

\title{LUNAR LASER RANGING TESTS OF\\ THE EQUIVALENCE PRINCIPLE WITH THE EARTH AND MOON}

\author{JAMES G. WILLIAMS, SLAVA G. TURYSHEV, DALE H. BOGGS}

\address{Jet Propulsion Laboratory, California Institute of  Technology,\\
4800 Oak Grove Drive, Pasadena, CA 91109, USA}

%\author{SECOND AUTHOR}
%
%\address{Group, Laboratory, Address\\
%City, State ZIP/Zone, Country\\
%second\_author@group.com}

\maketitle

\begin{history}
\received{Day Month Year}
\revised{Day Month Year}
\comby{Managing Editor}
\end{history}

\begin{abstract}
A primary objective of the Lunar Laser Ranging (LLR) experiment is to provide precise observations of the lunar orbit that contribute to a wide range of science investigations.  In particular, time series of the highly accurate measurements of the distance between the Earth and Moon provide unique information used to determine whether, in accordance with the Equivalence Principle (EP), both of these celestial bodies are falling towards the Sun at the same rate, despite their different masses, compositions, and gravitational self-energies.  35 years since their initiation, analyses of precision laser ranges to the Moon continue to provide increasingly stringent limits on any violation of the EP. Current LLR solutions give $(-1.0 \pm 1.4) \times 10^{-13}$ for any possible inequality in the ratios of the gravitational and inertial masses for the Earth and Moon, $\Delta(M_G/M_I)$.  This result, in combination with laboratory experiments on the weak equivalence principle, yields a strong equivalence principle (SEP) test of $\Delta(M_G/M_I)_{\tt SEP} = (-2.0 \pm 2.0) \times 10^{-13}$.  Such an accurate result allows other tests of gravitational theories. The result of the SEP test translates into a value for the corresponding SEP violation parameter $\eta$ of $(4.4 \pm 4.5)\times10^{-4}$, where $\eta = 4\beta -\gamma -3$ and both $\gamma$ and $\beta$ are parametrized post-Newtonian (PPN) parameters.  Using the recent result for the parameter $\gamma$ derived from the radiometric tracking data from the Cassini mission, the PPN parameter $\beta$  (quantifying the non-linearity of gravitational superposition) is determined to be $\beta - 1 = (1.2 \pm 1.1) \times 10^{-4}$.  We also present the history of the lunar laser ranging effort and describe the technique that is being used.  Focusing on the tests of the EP, we discuss the existing data, and characterize the modeling and data analysis techniques.  The robustness of the LLR solutions is demonstrated with several different approaches that are presented in the text.  We emphasize that near-term improvements in the LLR ranging accuracy will further advance the research of relativistic gravity in the solar system, and, most notably, will continue to provide highly accurate tests of the Equivalence Principle.
\end{abstract}


\section{Introduction}

The Equivalence Principle (EP) has been a focus of gravitational research for more than four hundred years.  Since the time of Galileo (1564-1642) it has been known that objects of different mass and composition accelerate at identical rates in the same gravitational field.  In 1602-04 through his study of inclined planes and pendulums, Galileo formulated a law of falling bodies that led to an early empirical version of the EP.  However, these famous results would not be published for another 35 years.  It took an additional fifty years before a theory of gravity that described these and other early gravitational experiments was published by Newton (1642-1727) in his Principia in 1687.  Newton concluded on the basis of his second law that the gravitational force was proportional to the mass of the body on which it acted, and by the third law, that the gravitational force is proportional to the mass of its source. 

Newton was aware that the \emph{inertial mass} $M_I$ appearing in the second law ${\bf F} = M_I {\bf a}$, might not be the same as the \emph{gravitational mass} $M_G$ relating force to gravitational field  ${\bf F} = M_G {\bf g}$. Indeed, after rearranging the two equations above we find ${\bf a} = (M_G/M_I){\bf g}$ and thus in principle materials with different values of the ratio $(M_G/M_I)$ could accelerate at different rates in the same gravitational field.  He went on testing this possibility with simple pendulums of the same length but different masses and compositions, but found no difference in their periods.  On this basis Newton concluded that $(M_G/M_I)$ was constant for all matter, and by a suitable choice of units the ratio could always be set to one, i.e. $(M_G/M_I) = 1$. Bessel (1784-1846) tested this ratio more accurately, and then in a definitive 1889 experiment E\"otv\"os was able to experimentally verify this equality of the inertial and gravitational masses to an accuracy of one part in $10^9$ (see Refs.~\refcite{Eotvos_1890,Eotvos_etal_1922,Bod_etal_1991}). 

Today, almost three hundred and twenty years after Newton proposed a comprehensive approach to studying the relation between the two masses of a body, this relation still continues to be the subject of modern theoretical and experimental investigations.  The question about the equality of inertial and passive gravitational masses arises in almost every theory of gravitation.  Nearly one hundred years ago, in 1915, the EP became a part of the foundation of Einstein's general theory of relativity; subsequently, many experimental efforts focused on testing the equivalence principle in the search for limits of general relativity.  Thus, the early tests of the EP were further improved by Roll et al.\cite{Roll_etal_1964} to one part in $10^{11}$. Most recently, a University of Washington group\cite{Baessler_etal_1999,Adelberger_2001} has improved upon Dicke's verification of the EP by several orders of magnitude, reporting $M_G/M_I - 1 < 1.4 \times 10^{-13}$.

The nature of gravity is fundamental to our understanding of our solar system, the galaxy and the structure and evolution of the universe.  This importance motivates various precision tests of gravity both in laboratories and in space.  To date, the experimental evidence for gravitational physics is in agreement with the general theory of relativity; however, there are a number of reasons to question the validity of this theory.  Despite the success of modern gauge field theories in describing the electromagnetic, weak, and strong interactions, it is still not understood how gravity should be described at the quantum level.  In theories that attempt to include gravity, new long-range forces can arise in addition to the Newtonian inverse-square law.  Even at the purely classical level, and assuming the validity of the equivalence principle, Einstein's theory does not provide the most general way to establish the space-time metric.  Regardless of whether the cosmological constant should be included, there are also important reasons to consider additional fields, especially scalar fields. 

Although scalar fields naturally appear in the modern theories, their inclusion predicts a non-Einsteinian behavior of gravitating systems.  These deviations from general relativity lead to a violation of the EP, modification of large-scale gravitational phenomena, and cast doubt upon the constancy of the ``constants.''  In particular, the recent work in scalar-tensor extensions of gravity that are consistent with present cosmological models\cite{Damour_Nordtvedt_1993a,Damour_Nordtvedt_1993b,Damour_etal_2002a,Damour_etal_2002b,Nordtvedt_2003,Turyshev_etal_2007,Turyshev_2008} predicts a violation of the EP at levels of $10^{-13}$ to $10^{-18}$. This prediction motivates new searches for very small deviations of relativistic gravity from general relativity and provides a robust theoretical paradigm and constructive guidance for further gravity experiments.  As a result, this theoretical progress has given a new strong motivation for high precision tests of relativistic gravity and especially those searching for a possible violation of the equivalence principle.  Moreover, because of the ever increasing practical significance of the general theory of relativity (i.e. its use in spacecraft navigation, time transfer, clock synchronization, standards of time, weight and length, etc) this fundamental theory must be tested to increasing accuracy.  

Today Lunar Laser Ranging (LLR) is well positioned to address the challenges presented above. The installation of the cornercube retroreflectors on the lunar surface more than 35 years ago with the Apollo 11 lunar landing, initiated a unique program of lunar laser ranging tests of the EP.  LLR provides a set of highly accurate distance measurements between an observatory on the Earth and a corner cube retroreflector on the Moon which is then used to determine whether, in accordance with the EP, these astronomical bodies are both falling towards the Sun at the same rate, despite their different masses and compositions.  These tests of the EP with LLR were among the science goals of the Apollo project.  Today this continuing legacy of the Apollo program\cite{Dickey_etal_1994} constitutes the longest running experiment from the Apollo era; it is also the longest on-going experiment in gravitational physics.  

Analyses of laser ranges to the Moon have provided increasingly stringent limits on any violation of the EP; they also enabled accurate determinations of a number of relativistic gravity parameters.  Ranges started in 1969 and have continued with a sequence of improvements for 35 years.  Data of the last decade are fit with an rms residual of 2~cm. This accuracy permits an EP test for the difference in the ratio of the gravitational and inertial masses for the Earth and Moon with uncertainty of $1.4 \times 10^{-13}$ (see Refs.~\refcite{Turyshev_etal_2004,Williams_Turyshev_Boggs_2004}).  The precise LLR data contribute to many areas of fundamental and gravitational physics, lunar science, astronomy, and geophysics.  With a new LLR station in progress and the possibility of new retro-reflectors on the Moon, lunar laser ranging remains on the front of gravitational physics research in the 21st century. 

This paper focuses on the tests of the EP with LLR.  To that extent, Section~\ref{sec:history}  discusses the LLR history, experimental technique, and the current state of the effort.  Section~\ref{sec:ep} is devoted to the discussion of the tests of the EP with the Moon. It also introduces various ``flavors'' of the EP and emphasizes the importance of the Earth and Moon as two test bodies to explore the Strong Equivalence Principle (SEP).  Section~\ref{sec:data} describes the existing LLR data including the statistics for the stations and reflectors, observational selection effects, and distributions.  Section~\ref{sec:model} introduces and characterizes the modeling and analysis techniques, focusing on the tests of the EP. In Section~\ref{sec:data_analysis} we discuss the details of the scientific data analysis using the LLR data set for tests of the EP. We present solutions for the EP and also examine the residuals in a search for any systematic signatures.  Section~\ref{sec:derived} focuses on the effects derived from the precision tests of the EP. Section~\ref{sec:ememrging_oops} introduces the near term emerging opportunities and addresses their critical role for the future progress in the tests of the equivalence principle with lunar laser ranging. We conclude with a summary and outlook.

\section{Lunar Laser Ranging: History and Techniques}
\label{sec:history} 

LLR accurately measures the round-trip time of flight for a laser pulse fired from an observatory on the Earth, bounced off of a corner cube retroreflector on the Moon, and returned to the observatory.  The currently available set of LLR measurements is more than 35 years long and it has become a major tool to conduct precision tests of the EP in the solar system.  Notably, if the EP were to be violated this would result in an inequality of gravitational and inertial masses and thus, it would lead to the Earth and the Moon falling towards the Sun at slightly different rates, thereby distorting the lunar orbit.  Thus, using the Earth and Moon as astronomical test bodies, the LLR experiment searches for an EP-violation-induced perturbation of the lunar orbit which could be detected with the available ranges. 

In this Section we discuss the history and current state for this unique experimental technique used to investigate relativistic gravity in the solar system.

\subsection{Lunar Laser Ranging History}
\label{sec:early_history}

The idea of using the orbit of the Moon to test foundations of general relativity belongs to R. H. Dicke, who in early 1950s suggested using powerful, pulsed searchlights on the Earth to illuminate corner retroreflectors on the Moon or a spacecraft.\cite{Alley_1972,Bender_etal_1973} The initial proposal was similar to what today is known as astrometric optical navigation which establishes an accurate trajectory of a spacecraft by photographing its position against the stellar background.  The progress in quantum optics that resulted in the invention of the laser introduced the possibility of ranging in early 1960s.  Lasers---with their spatial coherence, narrow spectral emission, small beam divergence, high power, and well-defined spatial modes---are highly useful for many space applications.  Precision laser ranging is an excellent example of such a practical use.  The technique of laser Q-switching enabled laser pulses of only a few nanoseconds in length, which allowed highly accurate optical laser ranging.  

%Finally, adoption of the well established methods of microwave ranging to optical techniques led to the remarkable precision and accuracy achieved with laser range measurements.  

Initially the methods of laser ranging to the Moon were analogous to radar ranging, with laser pulses bounced off of the lunar surface.  A number of these early lunar laser ranging experiments were performed in the early 1960's, both at the Massachusetts Institute of Technology and in the former Soviet Union at the Crimean astrophysics observatory.\cite{Abalakin-Kokurin-1981,Kokurin_2003}  However, these lunar surface ranging experiments were significantly affected by the rough lunar topography illuminated by the laser beam.  To overcome this difficulty, deployment of a compact corner retroreflector package on the lunar surface was proposed as a part of the unmanned, soft-landing Surveyor missions, a proposal that was never realized.\cite{Alley_1972}  It was in the late 1960's, with the beginning of the NASA Apollo missions, that the concept of laser ranging to a lunar corner-cube retroreflector array became a reality.  

\begin{figure}[!t]
    \begin{center} 
\begin{minipage}[t]{.46\linewidth}
 \epsfig{figure=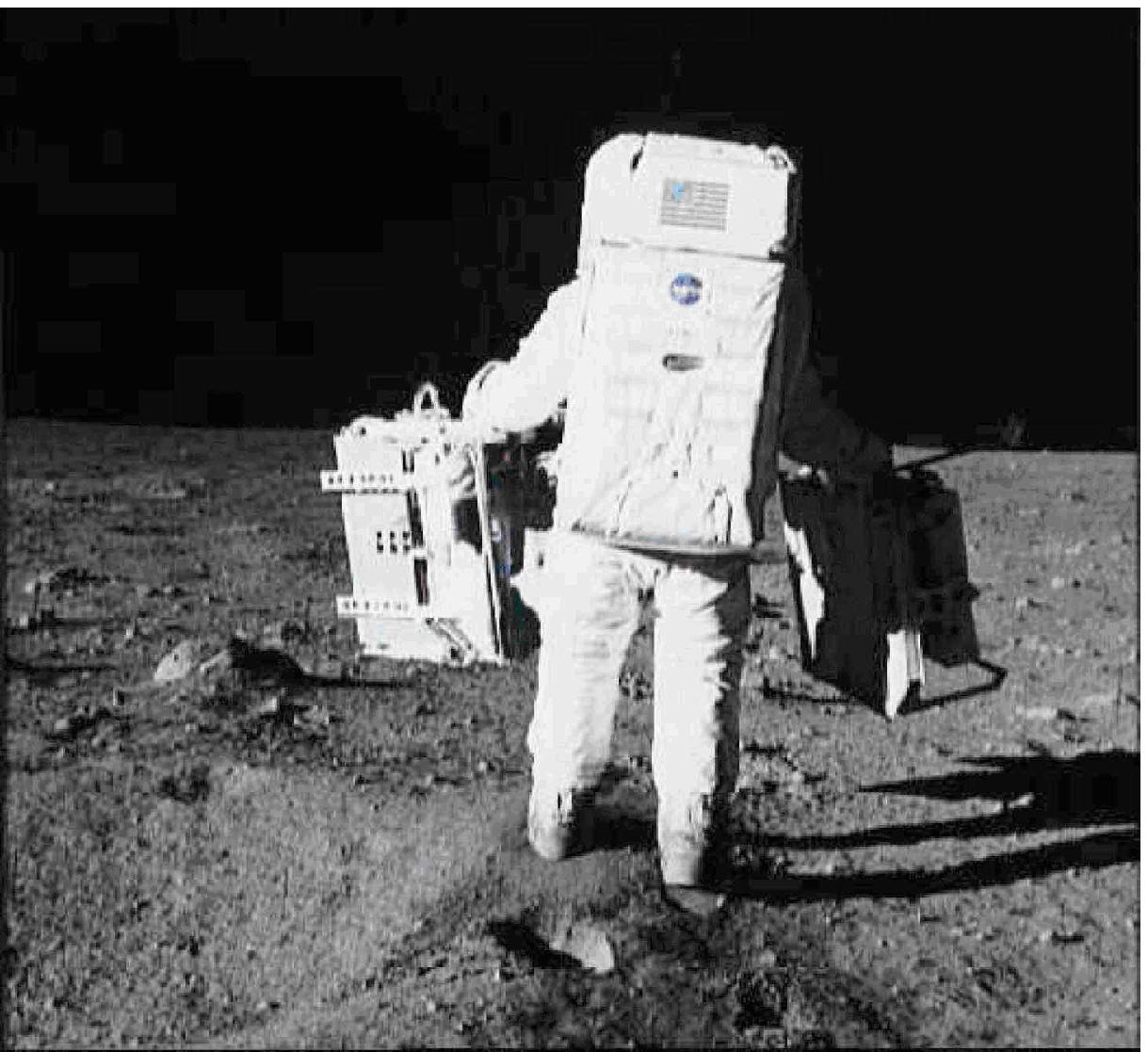,width=66mm} 
    \end{minipage}
\hskip 20pt
\begin{minipage}[t]{.46\linewidth} 
\epsfig{file=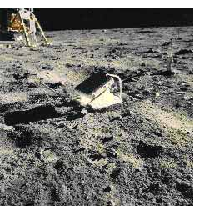,width=59mm}
    \end{minipage}
\caption{(a) The LLR retroreflector, at Buzz Aldrin's right side, being carried across the lunar surface by the Apollo 11 astronaut. (b) Apollo 11 laser retroreflector array.} \label{fig:1}
    \end{center}
\end{figure}

\begin{figure}[!t]
    \begin{center} 
\begin{minipage}[t]{.46\linewidth}
 \epsfig{figure=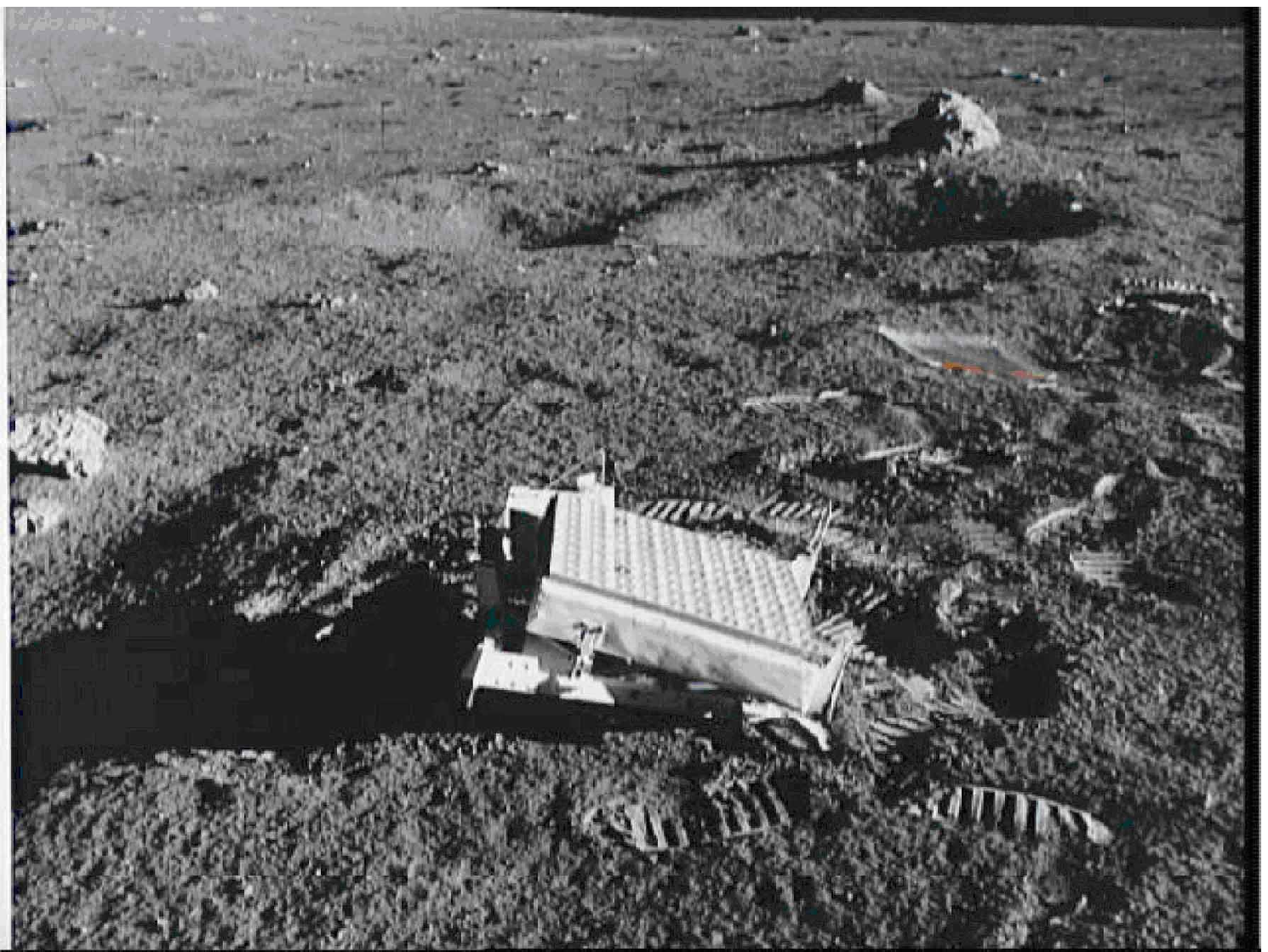,width=70mm} 
    \end{minipage}
\hskip 30pt
\begin{minipage}[t]{.44\linewidth} 
\epsfig{file=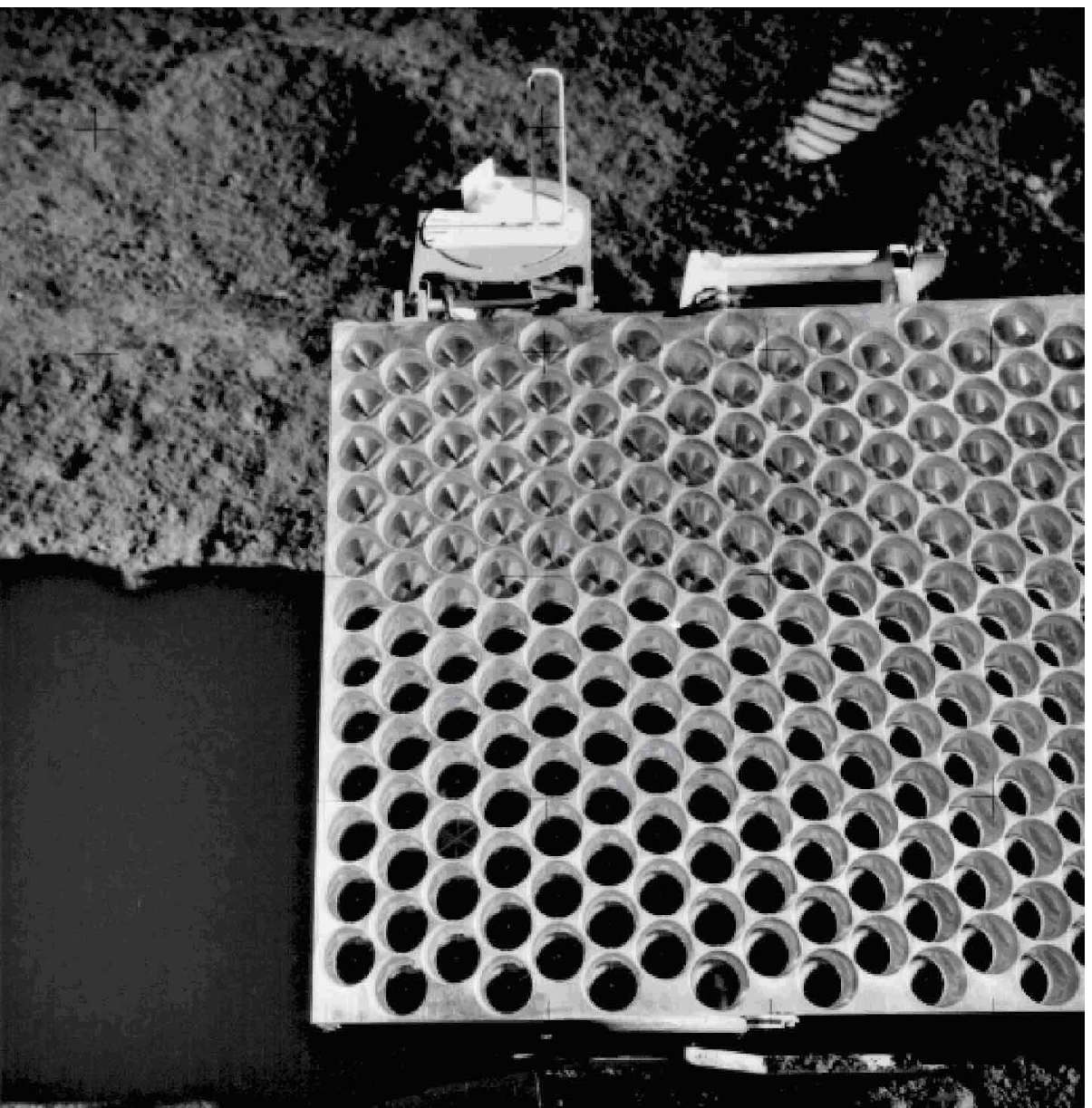,width=52mm}
    \end{minipage}
\caption{Apollo 14 (left) and Apollo 15 (right) LLR retroreflector arrays.} \label{fig:2}
    \end{center}
\end{figure}

\begin{figure}[!t]
    \begin{center} 
\begin{minipage}[t]{.46\linewidth}
 \epsfig{figure=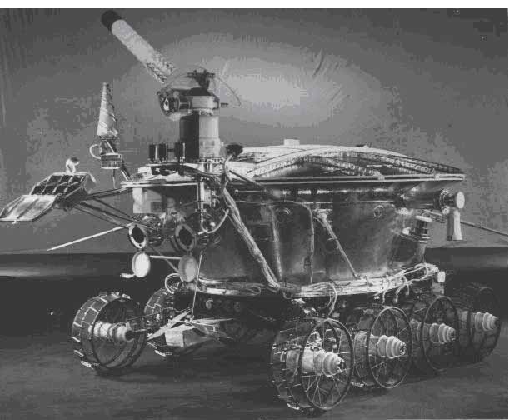,width=66mm} 
    \end{minipage}
\hskip 20pt
\begin{minipage}[t]{.46\linewidth} 
\epsfig{file=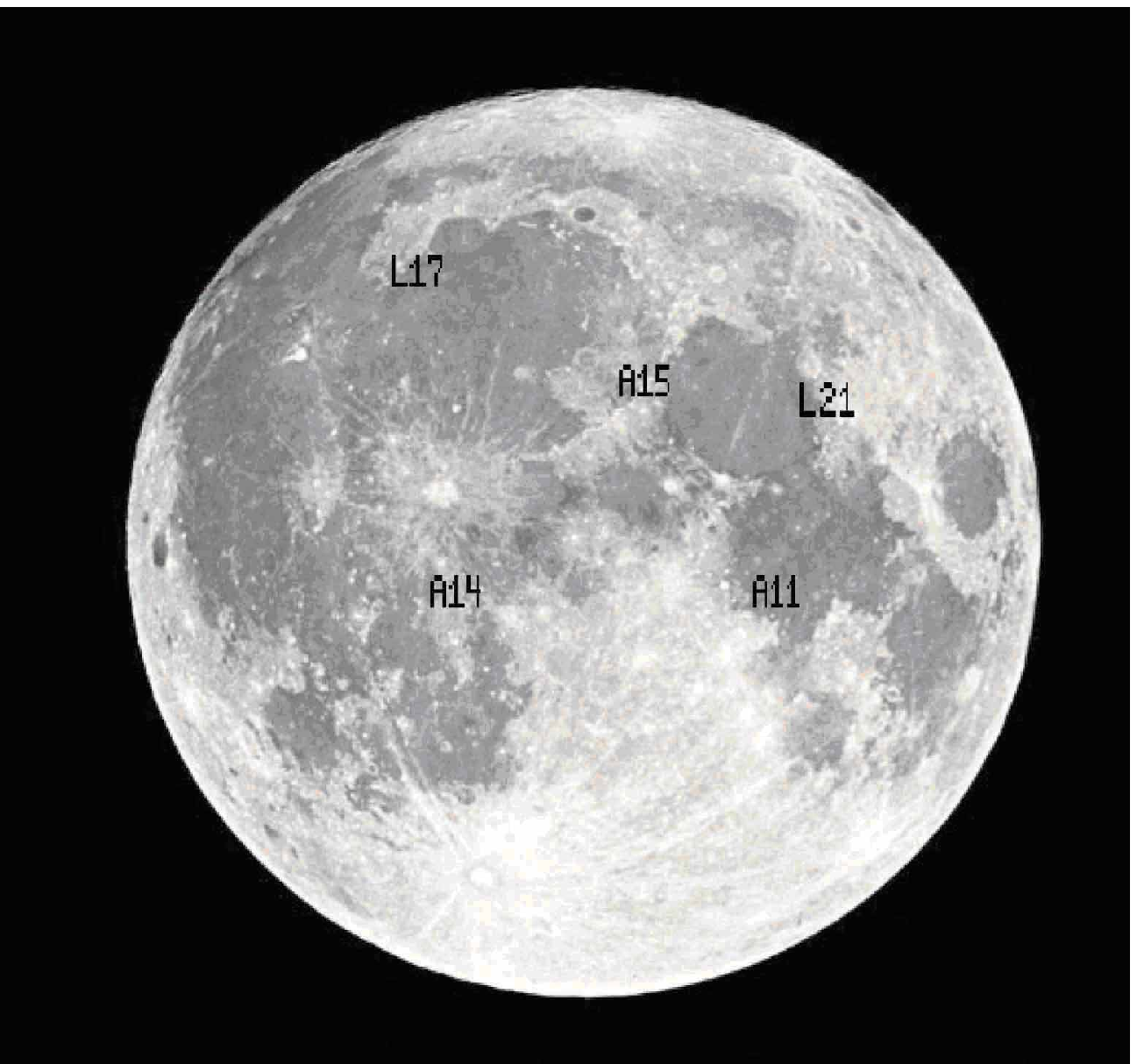,width=57mm}
    \end{minipage}
\caption{(a) Lunokhod 1 with the retroreflector array sticking out at far left. (b) The LLR retroreflector sites on the Moon.} \label{fig:3}
    \end{center}
\end{figure}

The scientific potential of lunar laser ranging led to the placement of retroreflector arrays on the lunar surface by the Apollo astronauts and the unmanned Soviet Luna missions to the Moon.  The first deployment of such a package on the lunar surface took place during the Apollo 11 mission (Figure~\ref{fig:1}) in the summer of 1969 and LLR became a reality\cite{Bender_etal_1973}. Additional retroreflector packages were set up on the lunar surface by the Apollo 14 and 15 astronauts (Figure~\ref{fig:2}). Two French-built retroreflector arrays were on the Lunokhod 1 and 2 rovers placed on the Moon by the Soviet Luna 17 and Luna 21 missions, respectively (Figure~\ref{fig:3}a). Figure~\ref{fig:3}b shows the LLR reflector sites on the Moon. 

The first successful lunar laser ranges to the Apollo 11 retroreflector were made with the 3.1~m telescope at Lick Observatory in northern California\footnote{The Lick Observatory website: {\tt http://www.ucolick.org/}}.\cite{Faller_etal_1969} The ranging system at Lick was designed solely for quick acquisition and confirmation, rather than for an extended program.  Ranges started at the McDonald Observatory in 1969 shortly after the Apollo 11 mission, while in the Soviet Union a sequence of laser ranges was made from the Crimean astrophysical observatory.\cite{Abalakin-Kokurin-1981,Kokurin_2003} A lunar laser ranging program has been carried out in Australia at the Orroral Observatory\footnote{The Orroral Observatory website: {\tt http://www.ga.gov.au/nmd/geodesy/slr/index.htm}}. Other lunar laser range detections were reported by the Air Force Cambridge Research Laboratories Lunar Ranging Observatory in Arizona\cite{AFCRL_1969}, the Pic du Midi Observatory in France\cite{Calame_etal_1970}, and the Tokyo Astronomical Observatory\cite{Kozai_1972}.

While some early efforts were brief and demonstrated capability, most of the scientific results came from long observing campaigns at several observatories.  The LLR effort at McDonald Observatory in Texas has been carried out from 1969 to the present.  The first sequence of observations was made from the 2.7~m telescope.  In 1985 ranging operations were moved to the McDonald Laser Ranging System (MLRS) and in 1988 the MLRS was moved to its present site\footnote{The McDonald Observatory website: {\tt http://www.csr.utexas.edu/mlrs/}}. The MLRS has the advantage of a shorter laser pulse and improved range accuracy over the earlier 2.7~m system, but the pulse energy and aperture are smaller.  From 1978 to 1980 a set of observations was made from Orroral in Australia.\cite{Luck_etal_1973,Morgan_King_1982}  Accurate observations began at the Observatoire de la C\^ote d'Azur (OCA) in 1984\footnote{The Observatoire de la C\^ote d'Azur website: {\tt http://www.obs-nice.fr/}} and continue to the present, though first detections were demonstrated earlier. Ranges were made from the Haleakala Observatory on the island of Maui in the Hawaiian chain from 1984 to 1990\footnote{The Haleakala Observatory website: {\tt http://koa.ifa.hawaii.edu/Lure/}}.  
 
%********************************
%********************************

%********************************
%********************************

%********************************
%********************************
 
Two modern stations which have demonstrated lunar capability are the Wettzell Laser Ranging System in Germany\footnote{The Wettzell Observatory website: {\tt http://www.wettzell.ifag.de/}} and the Matera Laser Ranging Station in Italy\footnote{The Matera Observatory: {\tt http://www.asi.it/html/eng/asicgs/geodynamics/mlro.html}}.  Neither is operational for LLR at present.  The Apache Point Observatory Lunar Laser ranging Operation (APOLLO) was recently built in New Mexico.\cite{Murphy_etal_2000,Williams_Turyshev_Murphy_2004,Murphy_etal_2007,Murphy_etal_2008,Turyshev_2008}

The two stations that have produced LLR observations routinely for decades are the McDonald Laser Ranging System (MLRS)\cite{Shelus_etal_2003}  in the United States and the OCA\cite{Veillet_etal_1993,Samain_etal_1998}  station in France.

\subsection{LLR and Fundamental Physics Today }
\label{sec:llr_funphys}
	
The analyses of LLR measurements contribute to a wide range of scientific disciplines, and are solely responsible for production of the lunar ephemeris. For a general review of LLR see Ref.~\refcite{Dickey_etal_1994}. An independent analysis for Ref.~\refcite{Chapront_etal_2002} gives geodetic and astronomical results. The interior, tidal response, and physical librations (rotational variations) of the Moon are all probed by LLR,\cite{Williams_etal_2001b,Williams_Dickey_2003} making it a valuable tool for lunar science.

\begin{figure}[!ht]
 \begin{center}
\noindent    
\epsfig{figure=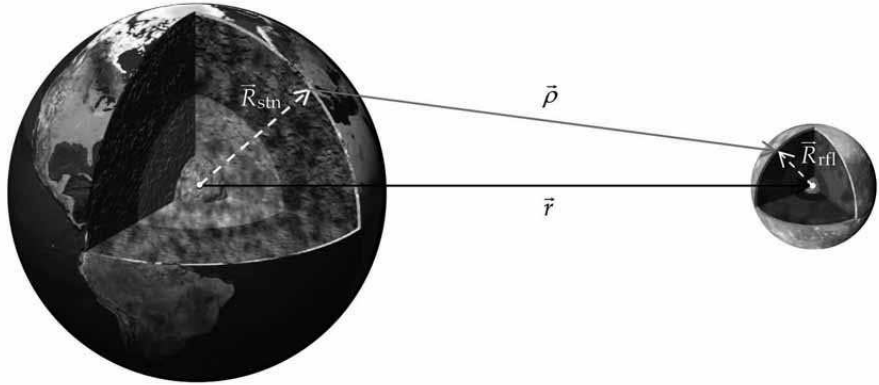,width=120mm}%,height=90mm}
\end{center}
\vskip -15pt 
  \caption{Lunar laser ranging accurately measures the distance between an observatory on Earth and a retroreflector on the Moon.  
 \label{fig:4}}
\end{figure}

The geometry of the Earth, Moon, and orbit is shown in Figure~\ref{fig:4}.  The mean distance of the Moon is 385,000 km, but there is considerable variation owing to the orbital eccentricity and perturbations due to Sun, planets, and the Earth's $J_2$ zonal harmonic.  The solar perturbations are thousands of kilometers in size and the lunar orbit departs significantly from an ellipse.  The sensitivity to the EP comes from the accurate knowledge of the lunar orbit.  The equatorial radii of the Earth and Moon are 6378 km and 1738 km, respectively, so that the lengths and relative orientations of the Earth-Moon vector, the station vector, and the retroreflector vector influence the range.  Thus, not only is there sensitivity of the range to anything which affects the orbit, there is also sensitivity to effects at the Earth and Moon. These various sensitivities allow the ranges to be analyzed to determine many scientific parameters.  

Concerning fundamental physics, LLR currently provides the most viable solar system technique for testing the Strong Equivalence Principle (SEP)--the statement that all forms of mass and energy contribute equivalent quantities of inertial and gravitational mass (see discussion in the following Section).  The SEP is more restrictive than the weak EP, which applies to non-gravitational mass-energy, effectively probing the compositional dependence of gravitational acceleration.   

%************
%**************

In addition to the SEP, LLR is capable of measuring the time variation of Newton's gravitational constant, $G$, providing the strongest limit available for the variability of this ``constant.'' LLR can also precisely measure the de Sitter precession--effectively a spin-orbit coupling affecting the lunar orbit in the frame co-moving with the Earth-Moon system's motion around the Sun.  The LLR results are also consistent with the existence of gravitomagnetism within 0.1\% of the predicted level\cite{Nordtvedt_1999,Nordtvedt_2003}; the lunar orbit is a unique laboratory for gravitational physics where each term in the parametrized post-Newtonian (PPN) relativistic equations of motion is verified to a very high accuracy. 

A comprehensive paper on tests of gravitational physics is Williams et al.\cite{Williams_etal_1996a}  A recent test of the EP is in Ref.~\refcite{Anderson_Williams_2001} and other general relativity tests are in Ref.~\refcite{Williams_etal_2002}.  An overview of the LLR gravitational physics tests is given by Nodtvedt.\cite{Nordtvedt_1999}  Reviews of various tests of relativity, including the contribution by LLR, are given in papers by Will.\cite{Will_1990,Will_2001}  Our recent paper, Ref.~\refcite{Williams_Turyshev_Murphy_2004}, describes the model improvements needed to achieve the mm-level accuracy for LLR. The most recent LLR results for gravitational physics are given in our recent paper of Ref.~\refcite{Williams_Turyshev_Boggs_2004}.

\section{Equivalence Principle and the Moon }
\label{sec:ep}

Since Newton, the question about equality of inertial and passive gravitational masses arises in almost every theory of gravitation.  Thus, almost one hundred years ago Einstein postulated that not only mechanical laws of motion, but also all non-gravitational laws should behave in freely falling frames as if gravity were absent.  If local gravitational physics is also independent of the more extended gravitational environment, we have what is known as the strong equivalence principle.  It is this principle that predicts identical accelerations of compositionally different objects in the same gravitational field, and also allows gravity to be viewed as a geometrical property of space-time--leading to the general relativistic interpretation of gravitation.  

The Equivalence Principle tests can therefore be viewed in two contexts: tests of the foundations of the standard model of gravity (i.e. general theory of relativity), or as searches for new physics because, as emphasized by Damour and colleagues,\cite{Damour_1996,Damour_2001} almost all extensions to the standard model of particle physics generically predict new forces that would show up as apparent violations of the EP.  The SEP became a foundation of Einstein's general theory of relativity proposed in 1915.  Presently, LLR is the most viable solar system technique for accurate tests of the SEP, providing stringent limits on any possible violation of general relativity - the modern standard theory of gravity.  

Below we shall discuss two different ``flavors'' of the Principle,  the weak and the strong forms of the EP that are currently tested in various experiments performed with laboratory tests masses and with bodies of astronomical sizes. 

\subsection{The Weak Form of the Equivalence Principle}
\label{sec:wep}

The weak form of the EP (the WEP) states that the gravitational properties of strong and electro-weak interactions obey the EP. In this case the relevant test-body differences are their fractional nuclear-binding differences, their neutron-to-proton ratios, their atomic charges, etc.  Furthermore, the equality of gravitational and inertial masses implies that different neutral massive test bodies will have the same free fall acceleration in an external gravitational field, and therefore in freely falling inertial frames the external gravitational field appears only in the form of a tidal interaction\cite{Singe_1960}. Apart from these tidal corrections, freely falling bodies behave as if external gravity were absent.\cite{Anderson_etal_1996}  General relativity and other metric theories of gravity assume that the WEP is exact.  However, extensions of the standard model of particle physics that contain new macroscopic-range quantum fields predict quantum exchange forces that generically violate the WEP because they couple to generalized ``charges'' rather than to mass/energy as does gravity.\cite{Damour_Polyakov_1994a,Damour_Polyakov_1994b}

In a laboratory, precise tests of the EP can be made by comparing the free fall accelerations, $a_1$ and $a_2$, of different test bodies.  When the bodies are at the same distance from the source of the gravity, the expression for the equivalence principle takes an elegant form:
{}
\begin{equation}
\frac{\Delta a}{a} = \frac{2(a_1- a_2)}{ a_1 + a_2} 
= \left(\frac{M_G}{M_I}\right)_1 -\left(\frac{M_G}{M_I}\right)_2  
= \Delta\left(\frac{M_G}{M_I}\right),		
\label{WEP_da}
\end{equation}

\noindent where $M_G$ and $M_I$ represent gravitational and inertial masses of each body. The sensitivity of the EP test is determined by the precision of the differential acceleration measurement divided by the degree to which the test bodies differ (e.g. composition).  

Since the early days of general relativity, Einstein's version of the Equivalence Principle became a primary focus of many experimental efforts.  Various experiments have been performed to measure the ratios of gravitational to inertial masses of bodies. Recent experiments on bodies of laboratory dimensions verify the WEP to a fractional precision $\Delta(M_G/M_I) \lesssim  10^{-11}$ by Roll et al.\cite{Roll_etal_1964}, to $\lesssim  10^{-12}$ by Refs.~\refcite{Braginsky_Panov_1972,Su_etal_1994} and more recently to a precision of $\lesssim 1.4\times 10^{-13}$ in Ref.~\refcite{Adelberger_2001}.  The accuracy of these experiments is sufficiently high to confirm that the strong, weak, and electromagnetic interactions each contribute equally to the passive gravitational and inertial masses of the laboratory bodies. 

This impressive evidence for laboratory bodies is incomplete for astronomical body scales. The experiments searching for WEP violations are conducted in laboratory environments that utilize test masses with negligible amounts of gravitational self-energy and therefore a large scale experiment is needed to test the postulated equality of gravitational self-energy contributions to the inertial and passive gravitational masses of the bodies\cite{Nordtvedt_1968a}.  Once the self-gravity of the test bodies is non-negligible (currently with bodies of astronomical sizes only), the corresponding experiment will be testing the ultimate version of the EP - the strong equivalence principle, that is discussed below. 

\subsection{The Strong Form of the Equivalence Principle}
\label{sec:sep}

In its strong form the EP is extended to cover the gravitational properties resulting from gravitational energy itself.  In other words, it is an assumption about the way that gravity begets gravity, i.e. about the non-linear property of gravitation. Although general relativity assumes that the SEP is exact, alternate metric theories of gravity such as those involving scalar fields, and other extensions of gravity theory, typically violate the SEP.\cite{Nordtvedt_1968a,Nordtvedt_1968b,Nordtvedt_1968c,Nordtvedt_1991} For the SEP case, the relevant test body differences are the fractional contributions to their masses by gravitational self-energy. Because of the extreme weakness of gravity, SEP test bodies that differ significantly must have astronomical sizes. Currently, the Earth-Moon-Sun system provides the best solar system arena for testing the SEP.

A wide class of metric theories of gravity are described by the parametrized post-Newtonian formalism,\cite{Nordtvedt_1968b,Will_1971,Will_Nordtvedt_1972} which allows one to describe within a common framework the motion of celestial bodies in external gravitational fields. Over the last 35 years, the PPN formalism has become a useful framework for testing the SEP for extended bodies.  To facilitate investigation of a possible violation of the SEP, in that formalism the ratio between gravitational and inertial masses, $M_G/M_I$, is expressed\cite{Nordtvedt_1968a,Nordtvedt_1968b} as
{}
\begin{equation}
\left[\frac{M_G}{M_I}\right]_{\tt SEP} = 1 + \eta\left(\frac{U}{Mc^2}\right),                               				\label{eq:MgMi}
\end{equation}

\noindent where $M$ is the mass of a body, $U$ is the body's gravitational self-energy $(U< 0)$, $Mc^2$ is its total mass-energy, and $\eta$ is a dimensionless constant for SEP violation.\cite{Nordtvedt_1968a,Nordtvedt_1968b,Nordtvedt_1968c}

Any SEP violation is quantified by the parameter $\eta$. In fully-conservative, Lorentz-invariant theories of gravity\cite{Will_1993,Will_2001} the SEP parameter is related to the PPN parameters by 
{}
\begin{equation}
\eta = 4\beta - \gamma -3.
\label{eq:eta}
\end{equation}

\noindent In general relativity $\beta = 1$ and $\gamma = 1$, 
so that $\eta = 0$. 

The self energy of a body B is given by 
\begin{equation}
\left(\frac{U}{Mc^2}\right)_B 
= - \frac{G}{2 M_B c^2}\int_B d^3{\bf x} d^3{\bf y} 
\frac{\rho_B({\bf x})\rho_B({\bf y})}{| {\bf x} - {\bf y}|}.        
\label{eq:omega}
\end{equation}

\noindent For a sphere with a radius $R$ and uniform density, $U/Mc^2 = -3GM/5Rc^2 = -3 v_E^2/10 c^2$, where $v_E$ is the escape velocity.  Accurate evaluation for solar system bodies requires numerical integration of the expression of Eq.~(\ref{eq:omega}). Evaluating the standard solar model\cite{Ulrich_1982} results in $(U/Mc^2)_S \sim -3.52 \times 10^{-6}$.  Because gravitational self-energy is proportional to $M^2$ (i.e. $U/Mc^2 \sim M$) and also because of the extreme weakness of gravity, the typical values for the ratio $(U/Mc^2)$ are $\sim 10^{-25}$ for bodies of laboratory sizes.  Therefore, the experimental accuracy of a part in $10^{13}$ (see Ref.~\refcite{Adelberger_2001}) which is so useful for the WEP is not a useful test of how gravitational self-energy contributes to the inertial and gravitational masses of small bodies.  To test the SEP one must utilize planetary-sized extended bodies where the ratio Eq.~(\ref{eq:omega}) is considerably higher.  

Nordtvedt\cite{Nordtvedt_1968a,Nordtvedt_1968c,Nordtvedt_1970} suggested several solar system experiments for testing the SEP.  One of these was the lunar test.  Another, a search for the SEP effect in the motion of the Trojan asteroids, was carried out by Orellana and Vucetich.\cite{Orellana_Vucetich_1988,Orellana_Vucetich_1993} Interplanetary spacecraft tests have been considered by Anderson et al.\cite{Anderson_etal_1996} and discussed by Anderson and Williams.\cite{Anderson_Williams_2001}  An experiment employing existing binary pulsar data has been proposed by Damour and Sch\"afer.\cite{Damour_Schafer_1991}  It was pointed out that binary pulsars may provide an excellent possibility for testing the SEP in the new regime of strong self-gravity\cite{Damour_Esposito-Farese_1996a,Damour_Esposito-Farese_1996b}, however the corresponding tests have yet to reach competitive accuracy\cite{Wex_2001,Lorimer_Freire_2004}. To date, the Earth-Moon-Sun system has provided the most accurate test of the SEP with LLR being the available technique.

\subsection{Equivalence Principle and the Earth-Moon system}
\label{sec:EP_earth_moon}

The Earth and Moon are large enough to have significant gravitational self energies and a lunar test of the equivalence principle was proposed by Nordtvedt.\cite{Nordtvedt_1968c}  Both bodies have differences in their compositions and self energies and the Sun provides the external gravitational acceleration.  For the Earth\cite{Flasar_Birch_1973,Williams_etal_1996a} a numerical evaluation of Eq.~(\ref{eq:omega}) yields:
{}
\begin{equation}
\left(\frac{U}{Mc^2}\right)_E = -4.64 \times 10^{-10}. 
\label{eq:earth}
\end{equation}

\noindent The two evaluations, with different Earth models, differ by only 0.1\%.  (A uniform Earth approximation is 10\% smaller in magnitude.)  A Moon model, with an iron core $\sim$20\% of its radius, gives 
{}
\begin{equation}
\left(\frac{U}{Mc^2}\right)_M = -1.90 \times 10^{-11}.   
\label{eq:moon }
\end{equation}

\noindent The subscripts $E$ and $M$ denote the Earth and Moon, respectively.  The lunar value is only 1\% different from the uniform density approximation which demonstrates its insensitivity to the model. The lunar value was truncated to two digits in Ref.~\refcite{Williams_etal_1996a}.  For the SEP effect on the Moon's position with respect to the Earth it is the difference of the two accelerations and self-energy values which is of interest.   
{}
\begin{equation}
\left(\frac{U}{Mc^2}\right)_E - \left(\frac{U}{Mc^2}\right)_M  = 
-4.45 \times 10^{-10}.
\label{eq:earth_moon}
\end{equation}

The Jet Propulsion Laboratory's (JPL) program which integrates the orbits of the Moon and planets considers accelerations due to Newtonian, geophysical and post-Newtonian effects.  Considering just the modification of the point mass Newtonian terms, the equivalence principle enters the acceleration $a_j$ of body $j$ as 
{}
\begin{equation}
{\bf a}_j = G \left(\frac{U}{Mc^2}\right)_j 
\sum_k M_k \frac{{\bf r}_{jk}}{r_{jk}^3},			        
\label{eq:accel}
\end{equation}

\noindent where ${\bf r}_{jk} = {\bf r}_k - {\bf r}_j$ is the vector from accelerated body $j$ to attracting body $k$ and $r_{jk} = |{\bf r}_{jk}|$.  For a more through discussion of the integration model see Ref.~\refcite{Standish_Williams_2005}.  

The dynamics of the three-body Sun-Earth-Moon system in the solar system barycentric inertial frame provides the main LLR sensitivity for a possible violation of the equivalence principle. In this frame, the quasi-Newtonian acceleration of the Moon with respect to the Earth, ${\bf a} = {\bf a}_M - {\bf a}_E$, is calculated to be:
{}
\begin{equation}
{\bf a} = - \mu^* \frac{{\bf r}_{EM}}{r^3_{EM}} 
- \left(\frac{M_G}{M_I}\right)_M \mu_S  \frac{{\bf r}_{SM}}{r^3_{SM}} + \left(\frac{M_G}{M_I}\right)_E \mu_S  \frac{{\bf r}_{SE}}{r^3_{SE}}, \label{eq:range1_m}
\end{equation}

\noindent where  $\mu^* = \mu_E (M_G/M_I)_M + \mu_M (M_G/M_I)_E$ and $\mu_k = G M_k$.  The first term on the right-hand side of Eq.(\ref{eq:range1_m}), is the acceleration between the Earth and Moon with the remaining pair being the tidal acceleration expression due to the solar gravity. The above acceleration is useful for either the weak or strong forms of the EP. 

For the SEP case, $\eta$ enters when expression Eq.~(\ref{eq:MgMi}) is is combined with  Eq.~(\ref{eq:range1_m}),
{}
\begin{equation}
{\bf a} = - \mu^* \frac{{\bf r}_{EM}}{r^3_{EM}} + \mu_S  \left[\frac{{\bf r}_{SE}}{r^3_{SE}} - \frac{{\bf r}_{SM}}{r^3_{SM}}\right] + \eta \mu_S \left[\left(\frac{U}{Mc^2}\right)_E \frac{{\bf r}_{SE}}{r^3_{SE}} - \left(\frac{U}{Mc^2}\right)_M \frac{{\bf r}_{SM}}{r^3_{SM}}\right]. 
\label{eq:range2_M}
\end{equation}

\noindent The presence of $\eta$ in $\mu^*$ modifies Kepler's third law to $n^2 a^3 = \mu^*$ for the relation between semimajor axis a and mean motion n in the elliptical orbit approximation. This term is notable, but in the LLR solutions $\mu_E + \mu_M$ is a solution parameter, or at least uncertain (see Sec.~\ref{sec:derived}), so this term does not provide a sensitive test of the equivalence principle, though its effect is implicit in the LLR solutions. The second term on the right-hand side with the differential acceleration toward the Sun is the Newtonian tidal acceleration. The third term involving the self energies gives the main sensitivity of the LLR test of the equivalence principle. Since the distance to the Sun is $\sim$390 times the distance between the Earth and Moon, the last term, is approximately $\eta$ times the difference in the self energies of the two bodies times the Sun's acceleration of the Earth-Moon center of mass. 

Treating the EP related tidal term as a perturbation Nordtvedt\cite{Nordtvedt_1968c} found a polarization of the Moon's orbit in the direction of the Sun with a radial perturbation 

\begin{equation}
\Delta r = S \left[\left(\frac{M_G}{M_I}\right)_E - \left(\frac{M_G}{M_I}\right)_M \right] \cos D,
\label{eq:range_r}
\end{equation}

\noindent where $S$ is a scaling factor of about $-2.9 \times 10^{13}$ mm (see Refs.~\refcite{Nordtvedt_1995,Damour_Vokrouhlicky_1996a,Damour_Vokrouhlicky_1996b}).  For the SEP, combining Eqs.~(\ref{eq:MgMi}) and (\ref{eq:range_r}) yields 
{}
\begin{eqnarray}
\Delta r &=& S \eta \left[\frac{U_E}{M_Ec^2}-\frac{U_M}{M_Mc^2} \right] \cos D, 	
\label{eq:S_cosD}\\
\Delta r &=& C_0 \eta \cos D. 
\label{eq:cosD}
\end{eqnarray}

\noindent Applying the difference in numerical values for self-energy for the Earth and Moon Eq.~(\ref{eq:earth_moon}) gives a value of $C_0$ of about 13~m (see Refs.~\refcite{Nordtvedt_etal_1995,Damour_Vokrouhlicky_1996a,Damour_Vokrouhlicky_1996b}). In general relativity $\eta = 0$. A unit value for $\eta$ would produce a displacement of the lunar orbit about the Earth, causing a 13~m monthly range modulation.  See subsection \ref{sec:EP_solution} for a comparison of the theoretical values of $S$ and $C_0$ with numerical results. This effect can be generalized to all similar three body situations. 

In essence, LLR tests of the EP compare the free-fall accelerations of the Earth and Moon toward the Sun. Lunar laser-ranging measures the time-of-flight of a laser pulse fired from an observatory on the Earth, bounced off of a retroreflector on the Moon, and returned to the observatory (see Refs.~\refcite{Dickey_etal_1994,Bender_etal_1973}). If the Equivalence Principle is violated, the lunar orbit will be displaced along the Earth-Sun line, producing a range signature having a 29.53 day synodic period (different from the lunar orbit period of 27 days). The first LLR tests of the EP were published in 1976 (see \refcite{Williams_etal_1976,Shapiro_etal_1976}). Since then the precision of the test has increased\cite{Dickey_Newhall_Williams_1989,Dickey_etal_1994,Chandler_etal_1994,Williams_etal_1996a,Williams_etal_1996b,Mueller_Nordtvedt_1998,Anderson_Williams_2001,Williams_etal_2002,Williams_Turyshev_Boggs_2004} until modern results are improved by two orders-of-magnitude. 

\subsection{Equivalence Principle and Acceleration by Dark Matter}
\label{sec:dark_matter}

At the scales of galaxies and larger there is evidence for unseen dark matter. Thus, observations of disk galaxies imply that the circular speeds are approximately independent of distance to the center of the galaxy at large distances. The standard explanation is that this is due to halos of unseen matter that makes up around 90\% of the total mass of the galaxies.\cite{Tremaine_1992} The same pattern repeats itself on larger and larger scales, until we reach the cosmic scales where a baryonic density compatible with successful big bang nucleosynthesis is less than 10\% of the density predicted by inflation, i.e. the critical density. Braginsky et al.\cite{Braginsky_etal_1992,Braginsky_1994} have studied the effect of dark matter bound in the galaxy but unbound to the solar system. Such galactic dark matter would produce an anisotropy in the gravitational background of the solar system. 

A possible influence of dark matter on the Earth-Moon system has been considered by Nordtvedt\cite{Nordtvedt_1994}, who has pointed out that LLR can also test ordinary matter interacting with galactic dark matter.  It was suggested that LLR data can be used to set experimental limits to the density of dark matter in the solar system by studying its effect upon the motion of the Earth-Moon system.  The period of the range signature is the sidereal month, 27.32 days.  An anomalous acceleration of $10^{-15}$ m/s$^2$ would cause a 2.5 cm range perturbation.  At this period there are also signatures due to other solution parameters: one component of station location, obliquity, and orbital mean longitude.  These parameters are separable because they contribute at other periods as well, but they are complications to the dark matter test.  

In 1995, Nordtvedt M\"uller, and Soffel published an upper limit of $3 \times 10^{-16}$~m/s$^2$ for a possible differential acceleration in coupling of dark matter to the different compositions of Earth and Moon.  This represented a stronger constraint by a factor of 150 than was achieved by the laboratory experiments searching for differential cosmic acceleration rates between beryllium and copper and between beryllium and aluminum.\cite{Smith_etal_1993,Su_etal_1994,Baessler_etal_1999,Adelberger_2001}

\section{Data}
\label{sec:data}

The accuracy and span of the ranges limit the accuracy of fit parameters.  This section describes the data set that is used to perform tests of the Equivalence Principle with LLR. The data taking is a day-to-day operation at the McDonald Laser Ranging System (MLRS) and the Observatoire de la C\^ote d'Azur (OCA) stations.  

LLR has remained a viable experiment with fresh results over 35 years because the data accuracies have improved by an order of magnitude.  See Section 4.1 below for a discussion and illustration (Figure~\ref{fig:5}) of that improvement.  The International Laser Ranging Service (ILRS)\footnote{International Laser Ranging Service (ILRS) website at {\tt http://ilrs.gsfc.nasa.gov/index.html}} provides lunar laser ranging data and their related products to support geodetic and geophysical research activities. 

\subsection{Station and Reflector Statistics}
\label{sec:stations}

LLR data have been acquired from 1969 to the present.  Each measurement is the round-trip travel time of a laser pulse between a terrestrial observatory and one of four corner cube retroreflectors on the lunar surface.  A normal point is the standard form of an LLR datum used in the analysis.  It is the result of a statistical combining of the observed transit times of several individual photons detected by the observing instrument within a relatively short time, typically a few minutes to a few tens of minutes.  

The currently operating LLR stations, McDonald Laser Ranging System in Texas\cite{Shelus_etal_2003} and Observatoire de la C\^ote d'Azur\cite{Samain_etal_1998}, typically detect 0.01 return photons per pulse during normal operation. A typical ``normal point'' is constructed from 3-100 return photons, spanning 10-45 minutes of observation.\cite{Dickey_etal_1994}

\begin{figure}[!t]
 \begin{center}
\noindent    
\epsfig{figure=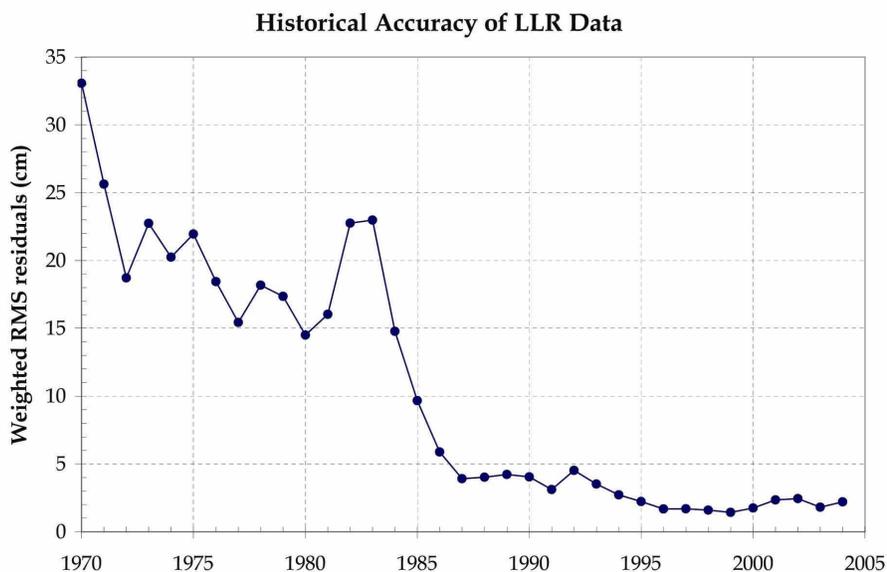,width=120mm}%,height=90mm}
\end{center}
\vskip -10pt 
  \caption{Annual rms residuals of LLR data from 1970 to 2004.  
 \label{fig:5}}
\end{figure}

The LLR data set for analysis has observations from McDonald Observatory, Haleakala Observatory, and OCA.  Figure~\ref{fig:5} shows the weighted RMS residual for each year.  Early accuracies using the McDonald Observatory's 2.7~m telescope hovered around 25 cm.  Equipment improvements decreased the ranging uncertainty to $\sim$15~cm later in the 1970s.  In 1985 the 2.7~m ranging system was replaced with the MLRS.  In the 1980s lunar ranges were also received from Haleakala Observatory on the island of Maui, Hawaii, and OCA in France.  Haleakala ceased lunar operations in 1990.  A sequence of technical improvements decreased the rms residual to the current $\sim$2~cm of the past decade.  The 2.7~m telescope had a greater light gathering capability than the newer smaller aperture systems, but the newer systems fired more frequently and had a much improved range accuracy.  The new systems do not distinguish returning photons against the bright background near full Moon, which the 2.7~m telescope could do.  There are some modern eclipse observations.  

%************
%**************

The first LLR test of the EP used 1523 normal points up to May 1975 with accuracies of 25 cm.  By April 2004, the data set has now grown to more than 15,554 normal points spanning 35 years, and the recent data is fit with $\sim$2~cm rms scatter.  Over time the post-fit rms residual has decreased due to improvements at both the McDonald and the OCA sites.  Averaged over the past four years there have been a total of several hundred normal points per year.  

The full LLR data set is dominated by three stations: the McDonald Station in Texas, the OCA station at Grasse, France, and the Haleakala station on Maui, Hawaii.  At present, routine ranges are being obtained only by the MLRS and OCA.  Figure~\ref{fig:3}b shows the distribution of the lunar retroreflectors.  Over the full data span 78\% of the ranges come from Apollo 15, 10\% from Apollo 11, 9\% from Apollo 14, 3\% from Lunokhod 2, and nothing from Lunokhod 1 (lost).

The notable improvement of the LLR data set with time implies comparable improvement in the determination of the solution parameters.  Data from multiple ranging sites to multiple retroreflectors are needed for a robust analysis effort. 

\subsection{Observational Influences and Selection Effects}
\label{sec:influences}

\begin{figure}[!t]
    \begin{center} 
\begin{minipage}[t]{.48\linewidth}
 \epsfig{figure=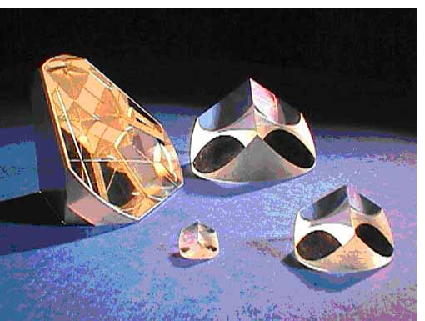,width=69mm} 
    \end{minipage}
\hskip 20pt
\begin{minipage}[t]{.45\linewidth} 
\epsfig{file=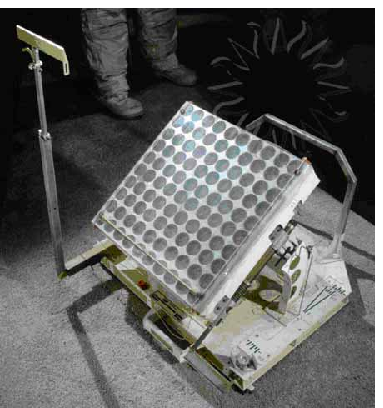,width=54mm}
    \end{minipage}
\caption{Corner-cube prisms are optical devices that return any incident light back in exactly the direction from which it came (left). An array of corner-cubes makes up the Apollo 11 lunar laser reflector (right).} \label{fig:6}
    \end{center}
\end{figure}

To range the Moon the observatories on the Earth fire a short laser pulse toward the target retroreflector array.  The outgoing laser beam is narrow and the illuminated spot on the Moon is a few kilometers across.  The retroreflectors are made up of arrays of corner cubes: 100 for Apollos 11 and 14, 300 for Apollo 15, and 14 for the Lunokhods.  At each corner cube (Figure~\ref{fig:6}) the laser beam enters the front face and bounces off of each of the three orthogonal faces at the rear of the corner cube.  The triply reflected pulse exits the front face and returns in a direction opposite to its approach.  The returning pulse illuminates an area around the observatory which is a few tens of kilometers in diameter.  The observatory has a very sensitive detector which records single photon arrivals.  Color and spatial filters are used to eliminate much of the background light.  Photons from different laser pulses have similar residuals with respect to the expected round-trip time of flight and are thus separated from the widely scattered randomly arriving background photons.  The resulting ``range'' normal point is the round trip light time for a particular firing time. (For more details on satellite and lunar laser ranging instrumentation, experimental set-up, and operations, consult papers by Degnan\cite{Degnan_1985,Degnan_1993,Degnan_2002} and Samain et al.\cite{Samain_etal_1998}

%********************************
%********************************

The signal returning from the Moon is so weak that single photons must be detected.  Not all ranging attempts are successful and the likelihood of success depends on the conditions of observation.  Observational effects may influence the strength of the signal, the background light which competes with the detection of the returning laser signal, the width of the outgoing or returning beam, and the telescope pointing.  Some of these observational influences select randomly and some select systematically, e.g. with phase of Moon, time of day, or time of year.  Selection with phase influences the equivalence principle test.  This subsection briefly discusses these observational influences and selection effects.  

The narrow laser beam must be accurately pointed at the target.  Seeing, a measure of the chaotic blurring of a point source during the transmission of light through the atmosphere, affects both the outgoing laser beam and the returning signal.  The beam's angular spread, typically a few seconds of arc ("), depends on atmospheric seeing so the spot size on the Moon is a few kilometers across at the Moon's distance (use 1.87 km/").  The amount of energy falling on the retroreflector array depends inversely on that spot area.  At the telescope's detector both a diaphragm restricting the field of view and a (few Angstrom) narrow-band color filter reduce background light.  When the background light is high the diaphragm should be small to reduce the interference and increase the signal-to-noise ratio.  When the seeing is poor the image size increases and this requires a larger diaphragm.  

The phase of the Moon determines whether a target retroreflector array is illuminated by sunlight or is in the dark.  These phase effects include the following influences. 

\begin{itemize}
\item[a)] The target illumination determines the amount of sunlight scattered back toward the observatory from the lunar surface near the target.  A sunlit surface increases the noise photons at the observatory's detector and decreases the signal to noise ratio.
  
\item[b)] The pointing technique depends on solar illumination around the target array.  Visual pointing is used when the target is sunlit while more difficult offset pointing, alignment using a displaced illuminated feature, is used when the target is dark.  

\item[c)] Retroreflector illumination by sunlight determines solar heating of the array and thermal effects on the retroreflector corner cubes.  A thermal gradient across a corner cube distorts the optical quality and spreads the return beam.  The Lunokhod corner cubes are about twice the size of the Apollo corner cubes and are thus more sensitive to thermal effects.  Also, the Lunokhod corner cubes have a reflecting coating on the three reflecting back sides while the Apollo corner cubes depend on total internal reflection.  The coating improves the reflected strength for beams that enter the front surface at an angle to the normal, where the Apollo efficiency decreases, but it also heats when sunlit.  Thus, the Lunokhod arrays have greater thermal sensitivity and are more difficult targets when heated by sunlight.  A retroreflector in the dark is in a favorable thermal environment, but the telescope pointing is more difficult.  
\end{itemize}

Whether the observatory is experiencing daylight or night determines whether sunlight is scattered toward the detector by the atmosphere.  As the Moon's phase approaches new, the fraction of time the Moon spends in the observatory's daylight sky increases while the maximum elevation of the Moon in the night sky decreases, so atmosphere-scattered sunlight is correlated with lunar phase.  

The beam returning from the Moon cannot be narrower than the diffraction pattern for a corner cube.  The diffraction pattern of a corner cube has a six-fold shape that depends on the six combinations of ways that light can bounce off of the three orthogonal reflecting faces.  
An approximate computation for green laser light (0.53 $\mu$m) gives 7 arcsec for the angular diameter of an Airy diffraction disk.  
%An approximate computation is made by substituting six 1.5 cm circles with the same total area as the 3.8 cm Apollo corner cubes.  Green laser light (0.53 $\mu$m) with 1.5 cm circles gives 17 arcsec for the angular diameter of an Airy diffraction disk.  
The larger Lunokhod corner cubes would give half that diffraction pattern size.  Thermal distortions, imperfections, and contaminating dust can make the size of the returning beam larger than the diffraction pattern.  So the returning spot size on the Earth is $\sim$30 km across for green laser light.  The power received by the telescope depends directly on the telescope's collecting area and inversely on the returning spot area.  Velocity-caused aberration of the returning beam is roughly 1" and is not a limitation since it is much smaller than the diffraction pattern.  

There are geometrical selection effects.  For the two operational northern ranging stations the Moon spends more time above the horizon when it is at northern declinations and less when south.  Also, atmospheric effects such as seeing and absorption increase at low elevation.  Consequently, there is selection by declination of the Moon.  This, along with climate, causes seasonal selection effects.  A station can only range the Moon when it is above the horizon which imposes selection at the 24 hr 50.47 min mean interval between meridian crossings.  

The best conditions for ranging occur with the Moon located high in a dark sky on a night of good seeing.  A daylight sky adds difficulty and full Moon is even more difficult.  A retroreflector in the dark benefits from not being heated by the Sun, but aiming the laser beam is more difficult.  New Moon occurs in the daylight sky near the Sun and ranging is not attempted since sensitive detectors are vulnerable to damage from bright light.  

\subsection{Data Distributions}
\label{sec:data_distribution}

\begin{figure}[!ht]
 \begin{center}
\noindent    
\epsfig{figure=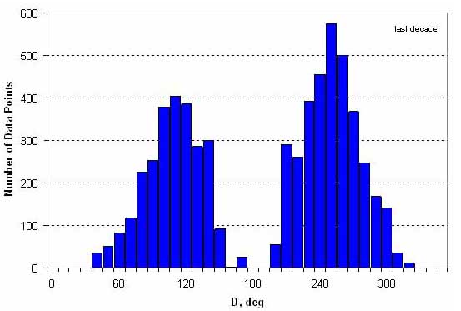,width=120mm}%,height=90mm}
\end{center}
\vskip -15pt 
%\caption{Distribution of last decade of data vs argument $D$, which has a 29.53 day period. \label{fig:7a}}
%\end{figure} 
%%**************
%%************
%\begin{figure}[!ht]
 \begin{center}
\noindent    \vskip -30pt 
\epsfig{figure=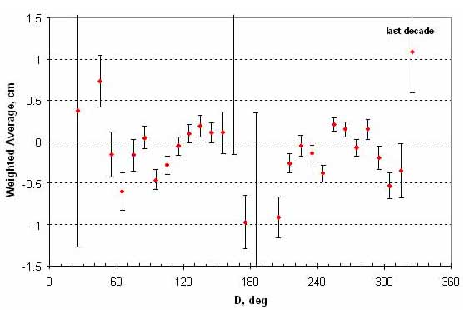,width=122mm}%,height=90mm}
\end{center}
\vskip -10pt 
  \caption{(a) Distribution of last decade of data vs argument $D$, which has a 29.53 day period. (b) Weighted average residual vs $D$ for last decade. \label{fig:7}}
%\caption{Weighted average residual vs $D$ for last decade.  
% \label{fig:7b}}
\end{figure}

\begin{figure}[!ht]
 \begin{center}
\noindent    
\epsfig{figure=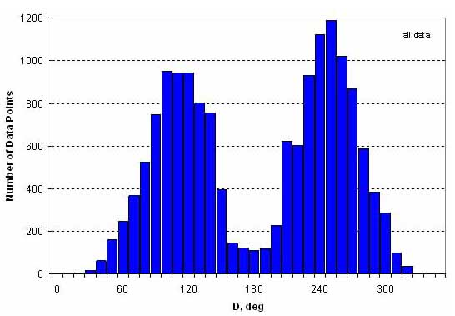,width=120mm}%,height=90mm}
\end{center}
\vskip -5pt 
%\caption{Distribution of all data vs argument $D$.  \label{fig:8a}}
%\end{figure} 
%%**************
%%************
%\begin{figure}[!ht]
%\begin{center}
%\noindent   
\vskip 0pt  \hskip 36pt 
\epsfig{figure=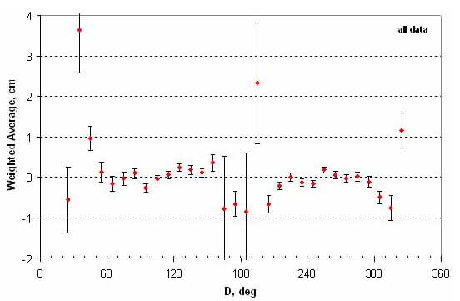,width=119mm}%,height=90mm}
%\end{center}
\vskip -10pt 
\caption{(a) Distribution of all data vs argument $D$. (b) Weighted average residual vs $D$ for all data. \label{fig:8}}
%
%\caption{Weighted average residual vs $D$ for all data. \label{fig:8b}}
\end{figure}

Observational selection effects shape the data distribution.  Several selection effects depend on the phase of the Moon and there is a dramatic influence on the distribution of the number of observations with phase.  The elongation of the Moon from the Sun is approximated with the angle $D$, the smooth polynomial representation of the difference in the mean longitudes for Sun and Moon.  Zero is near new Moon, 90$^\circ$ is near first quarter, 180$^\circ$  is near full Moon, and 270$^\circ$  is near last quarter.  Figure~\ref{fig:7}a illustrates the distribution of observations for the decade from 1995-2004 with respect to the angle $D$.  The shape of the curve results from the various selection effects discussed above.  There are no ranges near new Moon and few ranges near full Moon.  The currently operating observatories only attempt full Moon ranges during eclipses.  The original 2.7 m McDonald ranging system transmitted more energy in its longer pulse than currently operating systems, which gave it a higher single shot signal to noise ratio against a bright background.  It could range during full Moon as the distribution of the full data set for 1970-2004 shows in Figure~\ref{fig:8}a.  

\begin{figure}[!ht]
 \begin{center}
\noindent    
\epsfig{figure=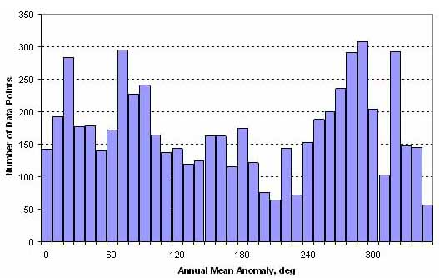,width=120mm}%,height=90mm}
\end{center}
\vskip -10pt 
%\caption{Annual mean anomaly distribution for the last decade of data.  \label{fig:9a}}
%\end{figure} 
%%**************
%%************
%\begin{figure}[!ht]
 \begin{center}
\noindent    \vskip -22pt 
\epsfig{figure=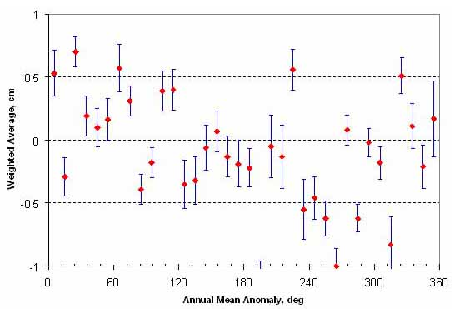,width=125mm}%,height=90mm}
\end{center}
\vskip -10pt 
\caption{(a) Annual mean anomaly distribution for the last decade of data. (b) Weighted average residual vs annual mean anomaly. 
\label{fig:9}}
%\caption{Weighted average residual vs annual mean anomaly.  \label{fig:9b}}
\end{figure}

Factors such as weather and the northern hemisphere location of the operating stations cause seasonal selection effects.  The distribution of the number of observations vs the mean anomaly of the Earth-Moon system about the Sun is shown in Figure~\ref{fig:9}a.  The annual mean anomaly is zero in the first week of January so that the mean anomaly is offset from calendar day of the year by only a few days.  There is considerable variation in the frequency of observation; the distribution is at its highest in fall and winter and at its lowest in summer.  

Other selection effects such as distance and declination also influence the data distribution and can be seen with appropriate histograms.  Nonuniform data distributions are one contribution to correlations between solution parameters.  

%************
%**************
 
%************
%************** 
 
%************
%************** 

\section{Modeling}
\label{sec:model}

Lunar Laser Ranging measures the range from an observatory on the Earth to a retroreflector on the Moon. The center-to-center distance of the Moon from the Earth, with mean value 385,000 km, is variable due to such things as orbit eccentricity, the attraction of the Sun, planets, and the Earth's bulge, and relativistic corrections. In addition to the lunar orbit, the range from an observatory on the Earth to a retroreflector on the Moon depends on the positions in space of the ranging observatory and the targeted lunar retroreflector. Thus, the orientation of the rotation axes and the rotation angles for both bodies are important.  Tidal distortions, plate motion, and relativistic transformations also come into play. To extract the scientific information of interest, it is necessary to accurately model a variety of effects. 

The sensitivity to the equivalence principle is through the orbital dynamics.  The successful analysis of LLR data requires attention to geophysical and rotational effects for the Earth and the Moon in addition to the orbital effects.  Modeling is central to the data analysis.  The existing model formulation, and its computational realization in computer code, is the product of much effort.  This section gives an overview of the elements included in the present model.

\subsection{Range Model}
\label{sec:range_model}

The time-of-flight (``range'') calculation consists of the round-trip ``light time'' from a ranging site on the Earth to a retroreflector on the Moon and back to the ranging site.  This time of flight is about 2.5 sec.  The vector equation for the one-way range vector $\boldsymbol{\rho}$ is 

\begin{equation}
\boldsymbol{\rho} = {\bf r} - {\bf R}_{\tt stn} + {\bf R}_{\tt rfl},
\label{eq:distance}
\end{equation}

\noindent where ${\bf r}$ is the vector from the center of the Earth to the center of the Moon, ${\bf R}_{\tt stn}$ is the vector from the center of the Earth to the ranging site, and ${\bf R}_{\tt rfl}$ is the vector from the center of the Moon to the retroreflector array (see Figure~\ref{fig:4} for more details).  The total time of flight is the sum of the transmit and receive paths plus delays due to atmosphere and relativistic gravitational delay 

\begin{equation}
t_3 - t_1 = (\rho_{12} + \rho_{23})/c + \Delta t_{\tt atm} + \Delta t_{\tt grav}. 
\label{eq:dt}
\end{equation}

\noindent The times at the Earth are transmit (1) and receive (3), while the bounce time (2) is at the Moon.  Due to the motion of the bodies the light-time computation is iterated for both the transmit and receive legs.  Since most effects effectively get doubled, it is convenient to think of 1 nsec in the round-trip time as being equivalent to 15 cm in the one-way distance.  

The center of mass of the solar system is treated as unaccelerated.  This solar system barycenter (SSB) is the coordinate frame for evaluating the above equations including relativistic computations.  First, the transmit time at the station is transformed to the SSB coordinate time (called $T_{\tt eph}$ by Standish\cite{Standish_1998} approximated by TDB), the basic computations are made in that SSB frame, and the computed receive time is transformed back to the station's time.  

\begin{equation}
(t_3 - t_1 )_{\tt stn} = t_3 - t_1 + \Delta t_{\tt trans}.
\label{eq:dtstn}
\end{equation}

The form of Eq.~(\ref{eq:distance}) separates the modeling problem into aspects related to the orbit, the Earth, and the Moon.  Eq.~(\ref{eq:dt}) shows that time delays must be added and Eq.~(\ref{eq:dtstn}) demonstrates modification of the round-trip-time-delay due to choice of reference frame.  For the discussion below we make a similar separation.  The dynamics of the orbits and lunar rotation come from a numerical integration, and those are the first two topics.  Earth and Moon related computations are discussed next.  The last topic is time delays and transformations.  

\subsubsection{Orbit Dynamics, ${\bf r}$} 

The lunar and planetary orbits and the lunar rotation result from a simultaneous numerical integration of the differential equations of motion.  The numerical integration model is detailed by Standish and Williams\cite{Standish_Williams_2005}.  Ephemerides of the Moon and planets plus lunar rotation are available at the Jet Propulsion Laboratory web site {\tt http://ssd.jpl.nasa.gov/}.  

The numerical integration of the motion of the Moon, planets, and Sun generates positions and velocities vs time.  The existing model for accelerations accounts for: 
\begin{itemize}
\item Newtonian and relativistic point mass gravitational interaction between the Sun, Moon, and nine planets. Input parameters include masses, orbit initial conditions, PPN parameters $\beta$ and $\gamma$, $\dot G$, and equivalence principle parameters $(M_G/M_I)$.  
\item Newtonian attraction of the largest asteroids.
\item Newtonian attraction between point mass bodies and bodies with gravitational harmonics: Earth ($J_2, J_3, J_4$), Moon (second- through fourth-degree spherical harmonics), and Sun ($J_2$).
\item Attraction from tides on both Earth and Moon includes both elastic and dissipative components.  There is a terrestrial Love number $k_2$ and a time delay for each of three frequency bands: semidiurnal, diurnal, and long period.  The Moon has a different Love number $k_2$ and time delay.  
\end{itemize}

\subsubsection{Lunar Rotation Dynamics}

The numerical integration of the rotation of the Moon generates three Euler angles and three angular velocities.  The torque model accounts for: 
\begin{itemize}
\item	Torques from the point mass attraction of Earth, Sun, Venus, Mars and Jupiter. The lunar gravity field includes second- through fourth-degree terms.
\item	Figure-figure torques between Earth ($J_2$) and Moon ($J_2$ and $C_{22}$).
\item	Torques from tides raised on the Moon include elastic and dissipative components. The formulation uses a lunar Love number $k_2$ and time delay.
\item	The fluid core of the Moon is considered to rotate separately from the mantle. A dissipative torque at the lunar solid-mantle/fluid-core interface couples the two\cite{Williams_etal_2001b}. There is a coupling parameter and the rotations of both mantle and core are integrated.
\item An oblate fluid-core/solid-mantle boundary generates a torque from the flow of the fluid along the boundary. This is a recent addition. 
\end{itemize}

\subsubsection{Effects at Earth, ${\bf R}_{\tt stn}$}
\begin{itemize}
\item The ranging station coordinates include rates for horizontal plate motion and vertical motion.
\item The solid-body tides are raised by Moon and Sun and tidal displacements on the Earth are scaled by terrestrial Love numbers $h_2$ and $l_2$.  There is also a core-flattening correction for a nearly diurnal term and a ``pole tide'' due to the time-varying part of the spin distortion.
\item	The orientation of the Earth's rotation axis includes precession and nutation. The body polar ($z$) axis is displaced from the rotation axis by polar motion.  The daily rotation includes UT1 variations.  A rotation matrix between the space and body frames incorporates these effects.  
\item	The motion of the Earth with respect to the solar system barycenter requires a Lorentz contraction for the position of the geocentric ranging station.
\end{itemize}
A compilation of Earth-related effects has been collected by McCarthy and Petit\cite{McCarthy_Petit_2003}.

\subsubsection{Effects at the Moon, ${\bf R}_{\tt rfl}$}

\begin{itemize}
\item	The Moon-centered coordinates of the retroreflectors are adjusted for solid-body tidal displacements on the Moon. Tides raised by Earth and Sun are scaled by the lunar displacement Love numbers $h_2$ and $l_2$.
\item	The rotation matrix between the space and lunar body frames depends on the three Euler angles that come from the numerical integration of the Euler equations.
\item	The motion of the Moon with respect to the solar system barycenter requires a Lorentz contraction for the position of the Moon-centered reflector.
\end{itemize}

\subsubsection{Time Delays and Transformations}

\begin{itemize}
\item	Atmospheric time delay $\Delta t_{\tt atm}$ follows Ref.~\refcite{Marini_Murray_1973}. It includes corrections for surface pressure, temperature and humidity which are measured at the ranging site.
\item	The relativistic time transformation has time-varying terms due to the motion of the Earth's center with respect to the solar system barycenter.  In addition, the displacement of the ranging station from the center of the Earth contributes to the time transformation.  The transformation changes during the $\sim$2.5 sec round-trip time and must be computed for both transmit and receive times.  
\item	The propagation of light in the gravity fields of the Sun and Earth causes a relativistic time delay $\Delta t_{\tt grav}$.
\end{itemize}

\subsubsection{Fit Parameters \& Partial Derivatives}

For each solution parameter in the least-squares fit there must be a partial derivative of the ``range'' with respect to that parameter. The partial derivatives may be separated into two types - geometrical and dynamical. 

Geometrical partials of range are explicit in the model for the time of flight.  Examples are partial derivatives of range with respect to geocentric ranging station coordinates, Moon-centered reflector coordinates, station rates due to plate motion, tidal displacement Love numbers $h_2$ and $l_2$ for Earth and Moon, selected nutation coefficients, diurnal and semidiurnal $UT1$ coefficients, angles and rates for the Earth's orientation in space, and ranging biases.

Dynamical partials of lunar orbit and rotation are with respect to parameters that enter into the model for numerical integration of the orbits and lunar rotation. Examples are dynamical partial derivatives with respect to the masses and orbit initial conditions for the Moon and planets, the masses of several asteroids, the initial conditions for the rotation of both the lunar mantle and fluid core, Earth and Moon tidal gravity parameters ($k_2$ and time delay), lunar moment of inertia combinations $(B-A)/C$ and $(C-A)/B$, lunar third-degree gravity field coefficients, a lunar core-mantle coupling parameter, equivalence principle $M_G/M_I$, PPN parameters $\beta$ and $\gamma$, geodetic precession, solar $J_2$, and a rate of change for the gravitational constant $G$. Dynamical partial derivatives for the lunar and planetary orbits and the lunar rotation are created by numerical integration. 

Considering Eqs.~(\ref{eq:distance}) and (\ref{eq:dt}), the partial derivative of the scalar range $\rho$ with respect to some parameter $p$ takes the form 

\begin{equation}
\frac{\partial \rho}{\partial {p}} = (\hat{\boldsymbol\rho} \cdot \frac{\partial {\boldsymbol\rho}}{\partial p}),
\label{eq:partial}
\end{equation}

\noindent where $\hat{\boldsymbol\rho} = {\boldsymbol\rho}/\rho$ is the unit vector.  From the three terms in Eq.~(\ref{eq:distance}), $\partial {\boldsymbol\rho}/\partial p$ depends on $\partial {\bf r}/\partial p$, $-\partial {\bf R}_{\tt stn}/\partial p$, and $\partial {\bf R}_{\tt rfl}/\partial p$.  

\begin{equation}
\frac{\partial \rho}{\partial {p}} 
= \left(\hat{\boldsymbol\rho} \cdot (\frac{\partial {\bf r}}{\partial p} - \frac{\partial {\bf R}_{\tt stn}}{\partial p} + \frac{\partial {\bf R}_{\tt rfl}}{\partial p} )\right).	
\label{eq:partialr}
\end{equation}

\noindent The dynamical partial derivatives of the orbit are represented by $ \partial {\bf r}/ \partial p$.  Rotation matrices are used to transform ${\bf R}_{\tt stn}$ and ${\bf R}_{\tt rfl}$ between body and space-oriented coordinates, so the partial derivatives of the rotation matrices depend on fit parameters involving the Earth and Moon Euler angles such as the Earth rotation, precession and nutation quantities and numerous lunar parameters which are sensitive through the rotation.  Only geometrical partials contribute to $\partial {\bf R}_{\tt stn}/\partial p$.  Both dynamical and geometrical partials affect $\partial {\bf R}_{\tt rfl}/\partial p$.  

\subsubsection{Computation}

The analytical model has its computational realization in a sequence of computer programs. Briefly these programs perform the following tasks. 
 
\begin{itemize}
\item[a)] Numerically integrate the lunar and planetary orbits along with lunar rotation. 
\item[b)] Numerically integrate the dynamical partial derivatives. 
\item[c)] Compute the model range for each data point, form the pre-fit residual, and compute range partial derivatives. At the time of the range calculation a file of integrated partial derivatives for orbits and lunar rotation with respect to dynamical solution parameters is read and converted to partial derivatives for range with respect to those parameters following Eq.~(\ref{eq:partialr}). The partial derivative for PPN $\gamma$ has both dynamical and geometrical components. 
\item[d)] Solve the weighted least-squares equations to get new values for the fit parameters. 
\item[e)] Generate and plot post-fit residuals. 
\end{itemize}

A variety of solutions can be made using different combinations of fit parameters.  Linear constraints between solution parameters can also be imposed.  The dynamical parameters from a solution can be used to start a new integration followed by new fits.  The highest quality ephemerides are produced by iterating the integration and fit cycle.  

\subsubsection{Data Weighting}

A range normal point is composed of from 3 to 100 single photon detections.  As the normal point comes from the station, the uncertainty depends on the calibration uncertainty and the time spread of the detected returned pulse.  The latter depends on the length of the outgoing laser pulse, spread at the lunar retroreflector due to tilt of the array, and detector uncertainty.  Gathering more photons reduces these return pulse length contributions to the normal point.  The analyst can also adjust the weightings according to experience with the residuals.  The analysis program includes uncertainty associated with the input UT1 and polar motion variations.  

\subsubsection{Solar Radiation Pressure}
\label{sec:solar_rad_press}

Solar radiation pressure, like the acceleration from an equivalence principle violation, is aligned with the direction from the Sun and it produces a perturbation with the 29.53 d synodic period.  Thus, this force on the Earth and Moon deserves special consideration for the most accurate tests of the equivalence principle.  This acceleration is not currently modeled in the JPL software.  Here we rely on the analysis of Vokrouhlicky\cite{Vokrouhlicky_1997} who considered incident and reflected radiation for both bodies plus thermal radiation from the Moon.  He finds a solar radiation perturbation of $-3.65 \pm 0.08$ mm $\cos D$ in the radial coordinate.  

\subsubsection{Thermal Expansion}

The peak to peak variation of surface temperature at low latitudes on the Moon is nearly 300$^\circ$C.  The lunar ``day'' is 29.53 days long.  This is the same period as the largest equivalence principle term so a systematic effect from thermal expansion is indicated. The phase of the thermal cycle depends on the retroreflector longitude.  

The Apollo retroreflector arrays and the Lunokhod 1 vehicle with the attached retroreflector array are shown in Figure~\ref{fig:1}-\ref{fig:3}.  The Apollo 11, 14, and 15 retroreflector arrays are close to the lunar surface and the center of each array front face is about 0.3, 0.2, and 0.3 m above the surface, respectively.  The Apollo corner cubes are mounted in an aluminum plate.  The thermal expansion coefficient for aluminum is about $2 \times 10^{-5}$/$^\circ$C.  If the Apollo arrays share the same temperature variations as the surface, then the total variation of thermal expansion will be 1 to 2 mm.  The Lunokhod 2 vehicle is 1.35 m high.  From images the retroreflector array appears to be just below the top and it is located in front of the main body of the Lunokhod.  We do not know the precise array position or the thermal expansion coefficient of the rover, but assuming the latter is in the range of $1 \times 10^{-5}$/$^\circ$C to $3 \times 10^{-5}$/$^\circ$C then the peak vertical thermal variation will be in the range of 3 to 10 mm.  The horizontal displacement from the center of the Lunokhod is poorly known, but it appears to be $\sim$1 m and the horizontal thermal variation will be similar in size to the vertical variation.  The thermal expansion cycle is not currently modeled.  For future analyses, it appears to be possible to model the thermal expansion of the Apollo arrays without solution parameters, but a solution parameter for the Lunokhod 2 thermal cycle expansion seems to be indicated.  

The soil is heated and subject to thermal expansion, but it is very insulating and the ``daily'' thermal variation decreases rapidly with depth.  So less displacement is expected from the thermal expansion of the soil than from the retroreflector array.  

\section{Data Analysis}
\label{sec:data_analysis}

This section presents analysis of the lunar laser ranging data to test the equivalence principle.  To check consistency, more than one solution is presented.  Solutions are made with two different equivalence principle parameters and different ways of establishing the masses of Earth and Moon.  Also, spectra of the residuals after fits, the post-fit residuals, are examined for systematics.  

\subsection{Solutions for Equivalence Principle}
\label{sec:EP_solution}

The solutions presented here use 15,554 ranges from March 1970 to April 2004.  The ranging stations include the McDonald Observatory, the Observatoire de la C\^ote d'Azur, and the Haleakala Observatory.  The ranges of the last decade are fit with a 2 cm weighted rms residual.  Planetary tracking data are used to adjust the orbits of the Earth and other planets in joint lunar and planetary fits.  The planetary data analysis does not include a solution parameter for the equivalence principle.  

Among the solution parameters are $GM_{\tt Earth+Moon}$, lunar orbit parameters including semimajor axis, Moon-centered retroreflector coordinates, geocentric ranging station coordinates, and lunar tidal displacement Love number $h_2$.  For additional fit parameters see the modeling discussion in Section \ref{sec:model}.  An equivalence principle violation can be solved for in two ways.  The first is a parameter for $M_G/M_I$ with a dynamical partial derivative generated from numerical integration.  The second solves for a coefficient of $\cos D$ in the lunar range, a one-term representation.  The latter approach was used in two papers\cite{Williams_etal_1976,Ferrari_etal_1980}, but the more sophisticated dynamical parameter is used in more recent JPL publications, namely Refs.~\refcite{Anderson_Williams_2001,Williams_Turyshev_Boggs_2004}.  Both approaches are exercised here to investigate consistency.  

Five equivalence principle solutions are presented in Table~\ref{tab:1} as EP 1 to EP 5.  Each of these solutions includes a standard set of Newtonian parameters in addition to one or more equivalence principle parameters.  In addition, the EP 0 solution is a comparison case which does not solve for an equivalence principle parameter.  The solution EP 1 solves for the $({M_G}/{M_I})$ parameter using the integrated partial derivative.  That parameter is converted to the coefficient of $\cos D$ in radial distance using the factor $S = -2.9 \times 10^{13}$ mm in Eq.~(\ref{eq:range_r}) from subsection \ref{sec:EP_earth_moon}.  The EP 2 case solves for coefficients of $\cos D$ and $\sin D$ in distance using a geometrical partial derivative.  Solution  EP 3 solves for the $({M_G}/{M_I})$ parameter along with coefficients of $\cos D$ and $\sin D$ in distance.  The EP 4 solution constrains the Sun/(Earth+Moon) and Earth/Moon mass ratios.  The EP 5 solution uses the mass constraints and also constrains the lunar $h_2$.  

%====================================
\begin{table}[t!]
%\begin{center}
\tbl{Five solutions for the equivalence principle.}  %\vskip 0pt
{\begin{tabular}{cccccc} \hline\hline
{\small Solution} & {\small $({M_G}/{M_I})$} & {\small conversion }
& {\small {\tt coef} $\cos D$ } & {\small {\tt coef} $\sin D$} & {\small Sun/(Earth+Moon)} \\
& {\small solution}, & {\small $ ({M_G}/{M_I})\rightarrow {\tt coef}$}   &{\small solution}, & {\small solution}, &\\
& $ \times \,10^{-13}$ &mm& mm & mm &\\\hline
EP 0 &&&&& $328900.5596 \pm 0.0011$\\
EP 1 &$ 0.30\pm1.42$ & $-0.9\pm4.1$ & &&	
$328900.5595 \pm 0.0012$\\
EP 2 &&& \hskip -12pt $-0.5 \pm 4.2$ & $0.9 \pm 2.1$ &
$328900.5596 \pm 0.0012$\\
EP 3 & $ 0.79\pm 6.09$ & ~$-2.3 \pm 17.7$ & $1.7 \pm 17.8$ & $0.9 \pm 2.1$ &	
$328900.5595 \pm 0.0012 $ \\
EP 4 & $ 0.21\pm1.30 $ & $-0.6\pm3.8$ &&	&
$328900.5597 \pm 0.0007$ \\
EP 5 & $ \hskip -10pt -0.11 \pm 1.30 $ & ~~$0.3 \pm 3.8$ &&&
$328900.5597 \pm 0.0007$ \\
\hline\hline
\end{tabular} \label{tab:1}}
%\end{center} 
\end{table}
%==========================================================  

The values in Table~\ref{tab:2} are corrected for the solar radiation pressure perturbation as computed by Vokrouhlicky.\cite{Vokrouhlicky_1997}  See the modeling subsection \ref{sec:solar_rad_press} on solar radiation pressure for a further discussion.  For the EP 3 case, with two equivalence principle parameters, the sum of the two $\cos D$ coefficients in Table~\ref{tab:1} is $-0.6\pm 4.2$ mm and that sum corrects to $3.1 \pm 4.2$ mm, which may be compared with the four entries in Table~\ref{tab:2}.

%====================================
\begin{table}[t!]
%\begin{center}
\tbl{Solutions for the equivalence principle corrected for solar radiation pressure.} %\vskip 0pt
{\begin{tabular}{ccccc} \hline\hline
Solution & $({M_G}/{M_I})$ & $({M_G}/{M_I})\rightarrow {\tt coef} $ 
& {\tt coef} $\cos D$ & {\tt coef} $\sin D$ \\
& solution & conversion & solution & solution \\
&&mm& mm & mm \\\hline
EP 1 & $(-0.96\pm 1.42) \times 10^{-13}$ & $2.8 \pm 4.1$ &&\\
EP 2 &&& $3.1 \pm 4.2$ & $0.9 \pm 2.1$\\
EP 4 & $(-1.05 \pm 1.30) \times 10^{-13}$ & $3.0 \pm 3.8$ && \\
EP 5 & $(-1.37 \pm 1.30) \times 10^{-13}$ & $4.0 \pm 3.8$ && \\
\hline\hline
\end{tabular} \label{tab:2}}
%\end{center} 
\end{table}
%========================================================== 

The equivalence principle solution parameters in Tables ~\ref{tab:1} and ~\ref{tab:2} are within their uncertainties for all cases except EP 5 in Table~\ref{tab:2}, and that value is just slightly larger.  Also, the EP 2 coefficient of $\cos D$ agrees reasonably for value and uncertainty with the conversion of the $M_G/M_I$ parameter of the EP 1, EP 4 and EP 5 solutions to a distance coefficient.  For the EP~3 solution, the sum of the converted $M_G/M_I$ coefficient and the $\cos D$ coefficient agrees with the other solutions in the two tables.  There is no evidence for a violation of the equivalence principle and solutions with different equivalence principle parameters are compatible.  

The difference in uncertainty between the $\sin D$ and $\cos D$ components of both the EP~2 and EP~3 solutions is due to the nonuniform distribution of observations with respect to $D$, as illustrated in Figures~\ref{fig:7}a and \ref{fig:8}a.  The $\sin D$ coefficient is well determined from observations near first and last quarter Moon, but the $\cos D$ coefficient is weakened by the decrease of data toward new and full Moon.  

The EP 3 case, solving for $M_G/M_I$ along with $\cos D$ and $\sin D$ coefficients, is instructive.  The correlation between the $M_G/M_I$ and $\cos D$ parameters is 0.972 so the two quantities are nearly equivalent, as expected.  The uncertainty for the two equivalence principle parameters increases by a factor of four in the joint solution, but the solution is not singular, so there is some ability to distinguish between the two formulations.  The integrated partial derivative implicitly includes terms at frequencies other than the $D$ argument (Nordtvedt, private communication, 1996) and it will also have some sensitivity to the equivalence principle influence on lunar orbital longitude.  The equivalence principle perturbation on lunar orbital longitude is about twice the size of the radial component and it depends on $\sin D$.  The ratio of Earth radius to lunar semimajor axis is $R_E/a \sim 1/60.3$, the parallax is about 1$^\circ$, so the longitude component projects into range at the few percent level.  

The uncertainties in the EP 3 solution can be used to check the theoretical computation of the coefficients $S$, which multiplies $\Delta (M_G/M_I)$, and $C_0$, which are associated with the $\cos D$ radial perturbation (subsection 3.3).  Given the high correlation, a first approximation of $S=-2.92 \times10^{13}$~mm is given by the ratio of uncertainties, and our knowledge that it must be negative.  A more sophisticated estimate of $S=-2.99 \times 10^{13}$~mm comes from computing the slope of the axis of the uncertainty ellipse for the two parameters.  Using expression Eq.~(\ref{eq:earth_moon}) for the difference in self energies of the Earth and Moon, the two preceding values give $\Delta r=13.0$~m $\eta \cos D$ and $\Delta r=13.3$~m $\eta \cos D$, respectively.  For comparison, the theoretical computations of Ref.~\refcite{Nordtvedt_1995} give $S=-2.9\times 10^{13}$~mm and $\Delta r=12.8$~m $\eta \cos D$, Damour and Vokrouhlicky\cite{Damour_Vokrouhlicky_1996a} give $S=-2.9427 \times 10^{13}$~mm, corresponding to $\Delta r=13.1$~m $\eta \cos D$, and Nordtvedt and Vokrouhlicky\cite{Nordtvedt_Vokrouhlicky_1997} give $S=-2.943\times 10^{13}$~mm and $\Delta r=13.1$~m $\eta \cos D$.  The numerical results here are consistent with the theoretical computations within a few percent.  

The EP 1 solution serves as an example for correlations.  The correlation of $M_G/M_I$ with both $GM_{\tt Earth+Moon}$ and osculating semimajor axis (at the 1969 epoch of the integration) is 0.46.  $GM$ and mean semimajor axis are connected through Kepler's third law given that the mean motion is very well determined.  The product of mean semimajor axis and mean eccentricity is well determined and the correlation of $M_G/M_I$ with osculating eccentricity is 0.45.  The correlation with the Earth-Moon mass ratio is 0.26.  

The value of $GM_{\tt Earth+Moon}$ is important for the equivalence principle solutions.  The Sun's $GM$ is defined in units of AU$^3$/day$^2$ so $GM_{\tt Earth+Moon}$ in those same units may be expressed as the mass ratio Sun/(Earth+Moon) as is done in Table~\ref{tab:1}.  The Sun/(Earth+Moon) mass ratio is a solution parameter in EP 0 through EP 3.  The solutions marked EP 4 and EP 5 use a value derived from sources other than LLR.  The Sun/(Earth+Moon) mass ratio is fixed at a value, with uncertainty, based on GM(Earth) from Ries et al.\cite{Ries_etal_1992} and an Earth/Moon mass ratio of $81.300570 \pm 0.000005$ from Konopliv et al.\cite{Konopliv_etal_2002}.  The uncertainty for $M_G/M_I$ is improved somewhat for solution EP 4.  With a fixed $GM$, the correlation with semimajor axis becomes small, as expected, but the correlation with the lunar $h_2$ is now $0.42$ and the $h_2$ solution value is $0.044 \pm 0.007$.  For comparison, solution EP 1 had a correlation of $-0.01$ and a solution value of $0.043 \pm 0.009$.  The solution EP 5 adds the lunar Love number $h_2$ to the constrained values using $h_2 = 0.0397$ from the model calculations of Williams et al.\cite{Williams_etal_2005}  A realistic model $h_2$ uncertainty is about 15\%, close to the EP 4 solution value, and the $M_G/M_I$ uncertainty is virtually the same as in the EP 4 solution.  All solutions use a model Love number $l_2$ value constrained to $0.0106$.  Considering the difficulty of precisely comparing uncertainties between analyses of different data sets, the gains for the last two constrained equivalence principle solutions are modest at best.  

Five solutions presented in this subsection have tested the equivalence principle.  They do not show evidence for a significant violation of the equivalence principle.  

\subsection{Spectra - Searching for Signatures in the Residuals}
\label{sec:spectra}

Part of the LLR data analysis is the examination of post-fit residuals including the calculation of overall and annual rms, a search for signatures at certain fundamental periods, and spectra over a spread of frequencies.  Direct examination of residuals can reveal some systematic effects but spectra of residuals, appropriately weighted for their uncertainties, can expose subtle effects.  

First consider the baseline solution EP 0 without an equivalence principle parameter.  The distribution of observations vs $D$ has been shown in the histograms of Figures~\ref{fig:7}a and \ref{fig:8}a.  The last decade of mean weighted residuals vs $D$ is presented in Figure~\ref{fig:7}b and all of the data is plotted in Figure~\ref{fig:8}b.  If an equivalence principle violation were present it would look like a cosine.  No such signature is obvious and a fit to the residuals gives a 1~mm amplitude, which is insignificant.  

\begin{figure}[!ht]
 \begin{center}
\noindent    
\epsfig{figure=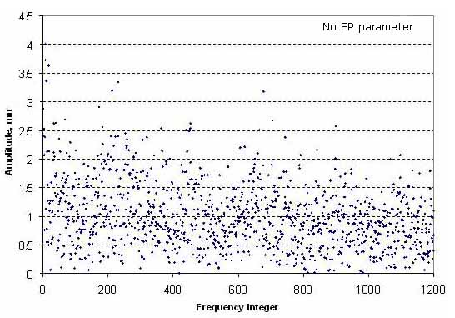,width=120mm}%,height=90mm}
\end{center}
\vskip -10pt 
%\caption{Spectrum of post-fit residuals without EP solution parameter.  Bin \#18 corresponds to 1 year, \#216 is synodic month, and \#239 is anomalistic month. \label{fig:10a}}
%\end{figure} 
%%**************
%%************
%\begin{figure}[!ht]
 \begin{center}
\noindent    \vskip -30pt 
\epsfig{figure=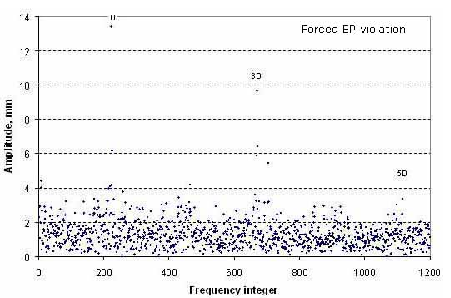,width=120mm}%,height=90mm}
\end{center}
\vskip -10pt 
\caption{(a) Spectrum of post-fit residuals without EP solution parameter. (b) Spectrum of residuals when a $\Delta (M_G/M_I)$ value of  $1.5\times 10^{-12}$ is forced into the solution. Frequency \#18 corresponds to 1 year, \#223 is synodic month, and \#239 is anomalistic month. 
 \label{fig:10}}
%\caption{Spectrum of residuals when a $\Delta (M_G/M_I)$ value of  $1.5\times 10^{-12}$ is forced into the solution. Bin \#18 corresponds to 1 year, \#216 is synodic month, and \#239 is anomalistic month. \label{fig:10b}}
\end{figure}

The LLR data are not evenly spaced or uniformly accurate so aliasing will be present in the spectra.  Here, a periodogram is computed by sequentially solving for sine and cosine components at equally spaced frequencies corresponding to periods from 18 years (6585~d) to 6~d.  Figure~\ref{fig:10}a shows the amplitude spectrum of the weighted post-fit residuals for the baseline solution.  Nothing is evident above the background at the 29.53 day synodic period (frequency \#223), which is consistent with the results of Table~\ref{tab:1}.  There are two notable features: a 3.6 mm peak at 1 yr and a broad increase at longer periods.  There are several uncompensated effects which might be contributing at 1 yr including loading effects on the Earth's surface height due to seasonal atmosphere and groundwater changes, and ``geocenter motion,'' the displacement of the solid body (and core) of the Earth with respect to the overall center of mass due to variable effects such as oceans, groundwater and atmosphere.  Averaged over more than 1000 frequencies the spectrum's background level is 1.2 mm.  Broad increases in the background near 1 month, $1/2$ month, and $1/3$ month etc, are due to aliasing.

%************
%************** 

For comparison, an equivalence principle signature was deliberately forced into another least-squares solution.  A finite $\Delta (M_G/M_I)$ value of $1.5\times 10^{-12}$, an order of magnitude larger than the uncertainty of the EP 1 solution of Tables~\ref{tab:1} and \ref{tab:2}, was constrained in a multiparameter least-squares solution.  The standard solution parameters were free to minimize the imposed equivalence principle signature as best they could.  Notably, $GM_{\tt Earth+Moon}$ and the Earth/Moon mass ratio were distorted from normal values by 5 and 3 times their realistic uncertainties, respectively, and the correlated orbit parameters also shifted by significant amounts.  The overall (35 year) weighted rms residual increased from 2.9 to 3.1 cm.  Figure~\ref{fig:10}b shows the spectrum of the residuals.  The two strongest spectral lines, 13 mm and 10 mm, are at the $D$ and $3D$ frequencies, respectively.  Detailed examination also shows weaker features, a $5D$ line and mixes of integer multiples of the $D$ frequency with the monthly and annual mean anomaly frequencies.  The expected equivalence principle signature of 44 mm $\cos D$ has been partly compensated by the least-squares adjustment of parameters for $GM$ and other quantities.  Note that the ratio of the 13 mm peak to the 1.2 mm background is compatible with the ratio of 44 mm (or $1.5 \times 10^{-12}$) to the equivalence principle uncertainty of 4.2 mm (or $1.4 \times 10^{-13}$) in Tables ~\ref{tab:1} and ~\ref{tab:2}.  The spectral amplitudes are computed one frequency at a time, but if the amplitudes of $\cos D$ and $\cos 3D$ are simultaneously fit to the post-fit residuals (not to the original ranges) then one gets $34 \cos D + 18 \cos 3D$ in mm.  This combination would be largest near new Moon, where there are no observations, and near full Moon, where there are very few accurate observations.  The spectrum for the baseline solution in Figure~\ref{fig:10}a shows no such lines.  In this figure the $\sim3$~mm peaks near 1 month and 1/3 month are at unassociated periods. 

In summary, a post-fit residual spectrum of baseline solution EP 0 without an equivalence principle parameter shows no evidence of any equivalence principle violation.  Manipulation shows that while a systematic equivalence principle signature can be diminished by adjusting other parameters during the least-squares solution, that compensation is only partly effective and a systematic effect cannot be eliminated.  It is also seen that the parameter uncertainties and correlations from the least-squares solutions are in reasonable agreement with the experience based on the spectra. 
 
\subsection{Classical Lunar Orbit}
\label{sec:lunar_orbit}

The JPL analyses use numerical integrations for the orbit and dynamical partial derivatives.  However, Keplerian elements and series expansions for the orbit give insight into the solution process.  

The Keplerian elements and mean distance of the Moon are summarized in Table ~\ref{tab:3}.  Note that the inclination is to the ecliptic plane, not the Earth's equator plane.  The lunar orbit plane precesses along a plane which is close to the ecliptic because solar perturbations are much more important than the Earth's $J_2$ perturbation.  A time average is indicated by $\left<...\right>$.  

%====================================
\begin{table}[t!]
%\begin{center}
\tbl{Lunar orbit.} %\vskip 3pt
{\begin{tabular}{rcc} \hline\hline
Mean distance 	& $\left<r \right>$ & 385,000.5 km \\
Semimajor axis	& ~~a = 1/$\left<1/r\right>$ ~~ & 384,399.0 km \\
Eccentricity	& $e$ &	0.0549 \\
Inclination to ecliptic plane	& $i$&	5.145$^\circ$ \\
\hline\hline
\end{tabular} \label{tab:3}}
%\end{center} 
\end{table}
%==========================================================	

Various lunar orbital angles and periods are summarized in Table ~\ref{tab:4}. These are mean angles represented by smooth polynomials.  The solar angles with annual periods are $l'$ for mean anomaly (the same as the mean anomaly of the Earth-Moon center of mass) and $L'$ for mean longitude (180$^\circ$ different from the mean longitude of the Earth-Moon center of mass). 

%====================================
\begin{table}[t!]
%\begin{center}
\tbl{Lunar angles.}% \vskip 3pt
{\begin{tabular}{rcc} \hline\hline
Angle & Symbol & Period \\\hline
Mean Longitude	& $L$ &	27.322 d \\
Mean Anomaly	& $l$ &	27.555 d \\
Mean Argument of Latitude & $F$ & 27.212 d \\
Mean Elongation of Moon from Sun & $D$ & 29.531 d \\
Mean Node & $\Omega$ &	18.61 yr \\
Mean Longitude of Perigee & $\varpi$ &	8.85 yr \\
Mean Argument of Perigee  & $\omega$ &	6.00 yr \\
\hline\hline
\end{tabular} \label{tab:4}}
%\end{center} 
\end{table}
%==========================================================

The lunar orbit is strongly perturbed by the Sun.  Chapront-Touz\'e and Chapront have developed an accurate series using computer techniques.  From that series (see \refcite{Chapront-Touze_Chapront_1988,Chapront-Touze_Chapront_1991}) a few large terms for the radial coordinate (in kilometers) are 
{}
\begin{eqnarray}
r &=& 385001 - 20905 \cos l - 3699 \cos(2D-l) - 2956 \cos 2D -\nonumber\\
&-& 570 \cos 2l + 246 \cos(2l-2D) + ... + 109 \cos D + ... 
\label{eq:rexp}
\end{eqnarray}

\noindent The constant first term on the right-hand side is the mean distance (somewhat larger than the semimajor axis), the $l$ and $2l$ terms are elliptical terms, and the remaining terms are from solar perturbations.  The amplitudes of the solar perturbation terms depend on the masses of the Earth, Moon, and Sun, as well as the lunar orbit and the Earth-Moon orbit about the Sun.  The periods of the periodic terms in the order given in Eq.~(\ref{eq:rexp}) are 27.555 d, 31.812 d, 14.765 d, 13.777 d, 205.9 d, and 29.531 d, so the different terms are well separated in frequency.  

If the equivalence principle is violated, there is a  dipole term in the expansion of the solar perturbation which gives the $\cos D$ term of subsection \ref{sec:EP_earth_moon}, see Refs.~\refcite{Nordtvedt_1968c,Nordtvedt_1995}.  When the equivalence principle is satisfied the dipole term has zero coefficient.  There is a classical $\cos D$ term which arises from the octupole ($P_3$) term in the expansion and that gives the 109 km amplitude in the series expansion for orbital $r$, Eq.~(\ref{eq:rexp}).  

The JPL Lunar Laser Ranging analyses use numerically integrated orbits, not series expansions (see \refcite{Chapront_etal_2002} for the polynomial expressions for lunar angles and an LLR data analysis with a higher reliance on analytical series).  The uncertainty of the solar perturbation corresponding to the classical $\cos D$ term is very small and is included in the final $M_G/M_I$ and amplitude uncertainties of the EP 1, EP 2, and EP 3 solutions of Tables ~\ref{tab:1} and ~\ref{tab:2}, since mass and orbit quantities are also solution parameters in those least-squares solutions.  

\subsection{Separation of the Equivalence Principle Signature}
\label{sec:separation}

The equivalence principle solution parameter, whether $M_G/M_I$ or $\cos D$, is significantly correlated with $GM$ of the Earth-Moon system and lunar semimajor axis.  The mean motion of the Moon is very well determined from the observations so Kepler's third law strongly relates the $GM$ and mean semimajor axis.  The correlation between $GM$ and $\cos D$ is related to the uneven distribution of observations for the angle $D$ (Figures~\ref{fig:7}a,\,\ref{fig:8}a).  The relation between the equivalence principle, $GM$ and the $D$ distribution has been extensively discussed by Nordtvedt\cite{Nordtvedt_1998}.  Some additional effects are briefly described by Anderson and Williams\cite{Anderson_Williams_2001}.  This subsection discusses the consequence of the $D$ distribution and other effects.  

The range may be derived from the vector Eq.~(\ref{eq:distance}).  The scalar range equation may be approximated as
{}
\begin{equation}
\rho \approx r - (\hat{\bf r}\cdot{\bf R}_{\tt stn}) + (\hat{\bf r}\cdot{\bf R}_{\tt rfl}) - ({\bf R}_{\tt stn} \cdot{\bf R}_{\tt rfl}) / r + ...			
\label{eq:(distexp)}
\end{equation}

\noindent The extended series expansion of the range equation is complicated, but with some consideration a few terms may be selected which are relevant to the equivalence principle solution.  The dot product between the orbital radius and the station vector involves large near daily and monthly variations and in solutions the station coordinates separate well from the other parameters.  Because of the good separation, this dot product will not be considered further here.  The series for orbital radius $r$ is given in Eq.~(\ref{eq:rexp}).  The mean distance for an elliptical orbit is given by $a(1+e^2/2)$, where for the Moon $a=384,399$ km and $e=0.0549$ (see Table \ref{tab:3}).  The terms with mean anomaly $l$ in the arguments have coefficients that depend on eccentricity.  The coefficient of the $2D$ term is scaled by the semimajor axis and while the coefficient has sensitivity to other parameters the semimajor axis scaling is the primary concern here.  The mean anomaly dependent terms have periods quite different from $D$ and will not be considered further.  Considering the dot product between reflector and orbit radius, the term of interest is $X u_1$, where $X$ is the reflector component toward the mean Earth direction, expressed in the body-referenced frame, and the expansion for the $x$ component of the unit vector from Moon center to Earth center in the same frame is Williams\cite{Williams_2005} 
{}
\begin{eqnarray}
u_1 &\approx& 0.99342 + 0.00337 \cos 2F + 0.00298 \cos 2l +\nonumber\\
&+& 0.00131 \cos 2D - 0.00124 \cos(2l-2D) + ...
\label{eq:u1}
\end{eqnarray}

\noindent The angle $F$ is the polynomial for the mean argument of latitude, the lunar angle measured from the node of the orbit on the ecliptic plane.  This angle is associated with the tilts of the orbit plane and the lunar equator plane to the ecliptic plane and it has a period of 27.212 days (Table~\ref{tab:4}).  

With the above considerations the relevant combination of terms is 

\begin{equation}
N \cos D + a ( 1.00157 - 0.00769 \cos 2D ) - X u_1 - ({\bf R}_{\tt stn}\cdot{\bf R}_{\tt rfl}) / r ,	
\label{eq:Nordt}
\end{equation}

\noindent where the first term represents an equivalence principle violation and $N, a,$ and $X$ are to be determined from the data.  The linear combination $1.0016 a - 0.9934 X$ is better determined by two orders-of-magnitude than either $a$ or $X$.  The separation of the different solution parameters is aided by the time variation of their multiplying functions in Eq.~(\ref{eq:Nordt}).  The periodic $2D$ term provides one way to separate $X$ and $a$.  If the angle $D$ were uniformly distributed, then the $D$ and $2D$ terms would be distinct.  The nonuniform distribution of $D$ (Figures~\ref{fig:7}a and \ref{fig:8}a) weakens the separation of the two periodicities and causes $N, a$ [and $GM_{\tt Earth+Moon}$], and $X$ to be correlated.  The separation of $X$ is aided by the periodic terms in $u_1$, such as the two half month terms with arguments $2F$ and $2l$, as well as the dot product between the station and reflector vectors, where $R_{\tt stn} / a = 1/60.3$ sets the scale for daily and longer period terms.  

A good equivalence principle test is aided by a) a good distribution of angle $D$, b) a good distribution of orbit angles $l$ and $F$, which is equivalent to a good distribution of orientations of the Moon's $x$ axis with respect to the direction to the Earth (optical librations), and c) a wide distribution of hour angles and declinations of the Moon as seen from the Earth.  Of these three, the first is the hardest for LLR to achieve for the reasons discussed in subsection \ref{sec:influences} on selection effects.  

\section{Derived Effects}
\label{sec:derived}

The solution EP 1 matches the EP test published in Ref.~\refcite{Williams_Turyshev_Boggs_2004}.  The data set of this paper has only one data point more than the data set of the published case.  Several consequences can be derived from the equivalence principle test including a test of the strong equivalence principle and PPN parameter $\beta$.  

\subsection{Gravity Shielding - the Majorana Effect}
\label{secLmajorana}

The possibility that matter can shield gravity is not predicted by modern theories of gravity, but it is a recurrent idea and it would cause a violation of the equivalence principle test.  Consequently, a brief discussion is given in this subsection.  

The idea of gravity shielding goes back at least as far as to the original paper by Majorana.\cite{Majorana_1920}  He proposed that the inverse square law of attraction should include an exponential factor $\exp(-h \int\rho(s) ds)$ which depends on the amount of mass between attracting mass elements and a universal constant $h$.  If mass shields gravity, then large bodies such as the Moon and Earth will partly shield their own gravitational attraction.  The observable ratio of gravitational mass to inertial mass would not be independent of mass, which would violate the equivalence principle.  Russell\cite{Russell_1921} realized that the large masses of the Earth, Moon and planets made the observations of the orbits of these bodies and the asteroid Eros a good test of such a possibility.  He made a rough estimate that the equivalence principle was satisfied to a few parts per million, which was much smaller than a numerical prediction based on Majorana's estimate for $h$.  

Majorana gave a closed form expression for a sphere's gravitational to inertial mass ratio.  For weak shielding a simpler expression is given by the linear expansion of the exponential term 

\begin{equation}
\frac{M_G}{M_I} \approx 1 - h f R \bar{\rho},
\label{eq:shield}
\end{equation}
where $f$ is a numerical factor, $\bar{\rho}$ is the mean density, and $R$ is the sphere's radius.  For a homogeneous sphere Majorana and Russell give $f=3/4$.  For a radial density distribution of the form $\rho(r)=\rho(0) (1-r^2/R^2)^n$ Russell\cite{Russell_1921} derives $f=(2n+3)^2/(12n+12)$.  

Eckhardt\cite{Eckhardt_1990} used an LLR test of the equivalence principle to set a modern limit on gravity shielding.  That result is updated as follows.  The uniform density approximation is sufficient for the Moon and $f R \bar{\rho} = 4.4 \times 10^8$ gm/cm$^2$.  For the Earth we use $n\approx 0.8$ with Russell's expression to get $f R\bar{\rho} = 3.4 \times10^9$ gm/cm$^2$.  Using the difference $-3.0\times10^9$ gm/cm$^2$ $h$ along with the LLR EP 1 solution from Table~\ref{tab:2} for the difference in gravitational to inertial mass ratios gives $h = (3 \pm 5) \times 10^{-23}$ cm$^2$/gm.  The value is not significant compared to the uncertainty.  To give a sense of scale to the uncertainty, for the gravitational attraction to be diminished by $1/2$ would require a column of matter with the density of water stretching at least half way from the solar system to the center of the galaxy.  The LLR equivalence principle tests give no evidence that mass shields gravity and the limits are very strong.  

\subsection{The Strong Equivalence Principle}
\label{sec:sep_solution}

The total equivalence principle results for the Earth-Moon system have been given in Table~\ref{tab:2}.  This test is a strong result in its own right.  The total equivalence principle is the sum of contributions from the WEP, which depends on composition, and the SEP, which depends on gravitational self energy.  This subsection extracts a result for the SEP by using WEP results from laboratory experiments at the University of Washington.  

Experiments by several groups have tested the WEP.  Several of these experiments with different test body compositions were compared in order to limit the WEP effect on the Earth-Moon pair to $10^{-12}$, see Refs.~\refcite{Adelberger_etal_1990a,Adelberger_etal_1990b}.  Recent laboratory investigations have synthesized the composition of the Earth and Moon\cite{Baessler_etal_1999,Adelberger_2001} by using test bodies which simulate the composition of core and mantle materials.  These WEP results are an order-of-magnitude more accurate.  

The most abundant element in the Earth is oxygen, followed by iron (30 weight \%), silicon and magnesium.\cite{Larimer_1986}  For the Moon, iron is in fourth place with about $1/3$ of the Earth's abundance.  The composition of the mantles of the Earth and Moon are similar, though there are differences (e.g. the Moon lacks the lower temperature volatiles such as water).  Iron and nickel are the heaviest elements which are abundant in both bodies.  Hence the difference in iron abundance, and associated siderophile elements, between the Earth and Moon is the compositional difference of most interest for the WEP.  

The Earth has a massive core ($\sim$1/3 by mass) with iron its major constituent and nickel and sulfur lesser components.  Several lines of evidence indicate that the Moon has a small core which is $<2$\% of its mass: moment of inertia\cite{Konopliv_etal_1998}, induced magnetic dipole moment\cite{Hood_etal_1999}, and rotational dynamics\cite{Williams_etal_2001b}.  The lunar core is presumed to be dominated by iron, probably alloyed with nickel and possibly sulfur, but the amount of information on the core is modest and evidence for composition is indirect.  In any case, most of the Fe in the Moon is in minerals in the thick mantle while for the Earth most of the Fe is in the metallic core.  For an example of lunar models see Ref.~\refcite{Kuskov_Kronrod_1998a,Kuskov_Kronrod_1998b}.  

For consideration of the WEP the iron content is the important difference in composition between the Earth and Moon.  Among the elements present at $>1$ weight \%, iron (and nickel for the Earth) have the largest atomic weights and numbers.  The two University of Washington test bodies reproduce the mean atomic weights and mean number of neutrons for the core material of the Earth (and probably the Moon) and both bodies' mantles.  

The Baessler et al.\cite{Baessler_etal_1999} and Adelberger\cite{Adelberger_2001} analyses use 38.2 \% for the fraction of mass of Fe/Ni core material in the whole Earth and 10.1 \% for the fraction in the Moon.  The difference in the experimental accelerations of the two test bodies is converted to the equivalent (WEP) difference in the acceleration of the Earth and Moon by multiplying by the difference (0.281).  Since the iron contents of the Earth and Moon are uncertain by a few percent, the effect of composition uncertainties is an order-of-magnitude less than the derived acceleration difference.  The Adelberger\cite{Adelberger_2001} result for the relative acceleration is given as $(1.0 \pm 1.4 \pm 0.2) \times 10^{-13}$, where the first uncertainty is for random errors and the second is for systematic errors.  We combine the systematic and random uncertainties and use
 
\begin{equation}
\left[\left(\frac{M_G}{M_I}\right)_E - \left(\frac{M_G}{M_I}\right)_M\right]_{\tt WEP} = (1.0 \pm 1.4) \times 10^{-13}. 	   
\label{eq:(WEP)}
\end{equation}

\noindent The strong equivalence principle test comes from combining solution EP 1 of Table~\ref{tab:2} with the above WEP result.  

\begin{equation}
\left[\left(\frac{M_G}{M_I}\right)_E - \left(\frac{M_G}{M_I}\right)_M\right]_{\tt SEP} = (-2.0 \pm 2.0) \times 10^{-13}.
\label{eq:(SEPLLR)}
\end{equation}

\noindent This combination of the LLR determination of the equivalence principle and the laboratory test of the weak equivalence principle provides the tightest constraint on the strong equivalence principle.  

\subsection{PPN Beta}
\label{sec:beta}

The test for a possible violation of the strong equivalence principle, the equivalence principle due to self-energy, is sensitive to a linear combination of PPN parameters.  For conservative theories this linear relation is $\eta = 4 \beta - \gamma - 3$, given by Eq.~(\ref{eq:eta}).  Using a good experimental determination of PPN $\gamma$, the SEP result can be converted into a result for PPN $\beta$.  

The test for any violation of the strong equivalence principle is sensitive to a linear combination of PPN quantities.  Considering only PPN $\beta$ and $\gamma$, divide the SEP determination of Eq.~(\ref{eq:(SEPLLR)}) by the numerical value from Eq. (\ref{eq:earth_moon}) to obtain 
{}
\begin{equation}
\eta = 4\beta - \gamma - 3 = (4.4\pm4.5)\times 10^{-4}.          \label{eq:etaLLR}
\end{equation}

\noindent This expression would be null for general relativity, hence the small value is consistent with Einstein's theory. 

The SEP relates to the non-linearity of gravity (how gravity affects itself), with the PPN parameter $\beta$ representing the degree of non-linearity. LLR provides great sensitivity to $\beta$, as suggested by the strong dependence of $\eta$ on $\beta$ in Eqs.~(\ref{eq:eta}) and (\ref{eq:etaLLR}). 

An accurate result for $\gamma$ has been determined by the Cassini spacecraft experiment.\cite{Bertotti_etal_2003}  Using high-accuracy Doppler measurements, the gravitational time delay allowed $\gamma$ to be determined to the very high accuracy of $\gamma - 1 = (2.1 \pm 2.3) \times 10^{-5}$.  This value of $\gamma$, in combination with $\eta$, leads to a significant improvement in the parameter $\beta$:  
{}
\begin{equation}
\beta - 1 = (1.2 \pm 1.1)\times 10^{-4}.                   
\label{eq:(betaLLR)}
\end{equation}

\noindent We do not consider this result to be a significant deviation of $\beta$ from unity. 

The PPN parameter $\beta$ has been determined by combining the LLR test of the equivalence principle, the laboratory results on the WEP, and the Cassini spacecraft determination of $\gamma$.  The uncertainty in $\beta$ is a dramatic improvement over earlier results.  The data set for the solutions in this chapter differs by only one point from that used in Ref.~\refcite{Williams_Turyshev_Boggs_2004}.  Consequently, the equivalence principle solution EP 1, and the derived result above for the strong equivalence principle, $\eta$ and $\beta$ are virtually the same as for the publication.  

\section{Emerging Opportunities}
\label{sec:ememrging_oops}

It is essential that the acquisition of new LLR data continue in the future.  Centimeter level accuracies are now achieved, and a further improvement is expected.  Analyzing improved data would allow a correspondingly more precise determination of gravitational physics and other parameters of interest.  In addition to the existing LLR capabilities, there are two near term possibilities that include the construction of the new LLR stations and development and deployment of either new sets of passive laser cornercube retroreflectors or active laser transponders pointed at Earth or both of these instruments.

In this Section we will discuss both of these emerging opportunities - the new LLR station in New Mexico and new LLR instruments on the Moon - for near term advancements in gravitational research in the solar system.

%also the emerging field of interplanetary laser ranging that recently had demonstrated its readiness for deployment in space.
%
%In this Section we will discuss both of these emerging opportunities - the new LLR station in New Mexico and interplanetary laser ranging technology - for near term advancements in gravitational research in the solar system. 

\subsection{New LLR Data and the APOLLO facility}
\label{sec:future_LLR}

LLR has remained a viable experiment with fresh results over 35 years because the data accuracies have improved by an order of magnitude (see Figure~\ref{fig:5}). A new LLR station should provide another order of magnitude improvement. The Apache Point Observatory Lunar Laser-ranging Operation (APOLLO) is a new LLR effort designed to achieve millimeter range precision and corresponding order-of-magnitude gains in measurements of fundamental physics parameters. Using a 3.5 m telescope the APOLLO facility will push LLR into the regime of stronger photon returns with each pulse, enabling millimeter range precision to be achieved.\cite{Murphy_etal_2000,Williams_Turyshev_Murphy_2004}  

An advantage that APOLLO has over current LLR operations is a 3.5 m astronomical quality telescope at a good site.  The site in southern New Mexico offers high altitude (2780 m) and very good atmospheric ``seeing'' and image quality, with a median image resolution of 1.1 arcseconds.  Both the image sharpness and large aperture combine to deliver more photons onto the lunar retroreflector and receive more of the photons returning from the reflectors, respectively.  Compared to current operations that receive, on average, fewer than 0.01 photons per pulse, APOLLO should be well into the multi-photon regime, with perhaps 1--10 return photons per pulse, depending on seeing.  With this signal rate, APOLLO will be efficient at finding and tracking the lunar signal, yielding hundreds of times more photons in an observation than current operations deliver. In addition to the significant reduction in random error (1/$\sqrt{N}$ reduction), the high signal rate will allow assessment and elimination of systematic errors in a way not currently possible. This station is designed to deliver lunar range data accurate to one millimeter. The APOLLO instrument started producing useful ranges in 2006, thereby, initiating the regular delivery of LLR data with much improved accuracy.\cite{Murphy_etal_2000,Williams_Turyshev_Murphy_2004,Murphy_etal_2007,Murphy_etal_2008}

The high accuracy LLR station installed at Apache Point should provide major opportunities (see Refs.~\refcite{Murphy_etal_2000,Williams_Turyshev_Murphy_2004,Murphy_etal_2008} for details).  The APOLLO project will push LLR into the regime of millimetric range precision which translates into an order-of-magnitude improvement in the determination of fundamental physics parameters.  An Apache Point 1 mm range accuracy corresponds to $3 \times 10^{-12}$ of the Earth-Moon distance.  The resulting LLR tests of gravitational physics would improve by an order of magnitude: the Equivalence Principle would give uncertainties approaching $10^{-14}$, tests of general relativity effects would be $<$0.1\%, and estimates of the relative change in the gravitational constant would be 0.1\% of the inverse age of the universe.  This last number is impressive considering that the expansion rate of the universe is approximately one part in $10^{10}$ per year.  Therefore, the gain in our ability to conduct even more precise tests of fundamental physics is enormous, thus this new instrument stimulates development of better and more accurate models for the LLR data analysis at a mm-level.   

\subsection{New retroreflectors and laser transponders on the Moon}
\label{sec:lunar-efforts}

%\subsection{Future Laser Ranging Opportunities}
%\label{sec:future_laser_ranging}

There are two critical factors that control the progress in the LLR-enabled science -- the distribution of retroreflectors on the lunar surface and their passive nature.  Thus, the four existing arrays\cite{Dickey_etal_1994} are distributed from the equator to mid-northern latitudes of the Moon and are placed with modest mutual separations relative to the lunar diameter.  Such a distribution is not optimal; it limits the sensitivity of the ongoing LLR science investigations.  The passive nature of reflectors causes signal attenuation proportional to the inverse 4th power of the distance traveled by a laser pulse.  The weak return signals drive the difficulty of the observational task; thus, only a handful of terrestrial SLR stations are capable of also carrying out the lunar measurements, currently possible at cm-level.

The intent to return to the Moon was announced in January 2004. NASA is planning to return to the Moon in 2009 with Lunar Reconnaissance Orbiter, and later with robotic landers, and then with astronauts in the next decade. The return to the Moon provides an excellent opportunity for LLR, particularly if additional LLR instruments will be placed on the lunar surface at more widely separated locations.  Due to their potential for new science investigations, these instruments are well justified.

\subsubsection{New retroreflector arrays}

Future ranging devices on the Moon might take two forms, namely passive retroreflectors and active transponders. The advantages of passive retroreflector arrays are their long life and simplicity. The disadvantages are the weak returned signal and the spread of the reflected pulse arising from lunar librations, which can change the retroreflector orientation up to 10 degrees with respect to the direction to the Earth. 

Range accuracy, data span, and distributions of earth stations and retroreflectors are important considerations for future LLR data analysis.  Improved range accuracy helps all solution parameters.  Data span is more important for some parameters, e.g. change in $G$, precession and station motion, than others. New retroreflectors optimized for pulse spread, signal strength, and thermal effects, will be valuable at any location on the moon.  

Overall, the separation of lunar 3-dimensional rotation, the rotation angle and orientation of the rotation axis (also called physical librations), and tidal displacements depends on a good geographical spread of retroreflector array positions.  The current three Apollo sites plus the infrequently observed Lunokhod 2 are close to the minimum configuration for separation of rotation and tides, so that unexpected effects might go unrecognized.  A wider spread of retroreflectors could improve the sensitivity to rotation/orientation angles and the dependent lunar science parameters by factors of up to 2.6 for longitude and up to 4 for pole orientation.  The present configuration of retroreflector array locations is quite poor for measuring lunar tidal displacements.  Tidal measurements would be very much improved by a retroreflector array near the center of the disk, longitude 0 and latitude 0, plus arrays further from the center than the Apollo sites.  

Lunar retroreflectors are the most basic instruments, for which no power is needed.  Deployment of new retroreflector arrays is very simple: deliver, unfold, point toward the Earth and walk away.  Retroreflectors should be placed far enough away from astronaut/moonbase activity that they will not get contaminated by dust.  One can think about the contribution of smaller retroreflector arrays for use on automated spacecraft and larger ones for manned missions.  One could also benefit from co-locating passive arrays and active transponders and use a few LLR capable stations ranging retroreflectors to calibrate the delay vs. temperature response of the transponders (with their more widely observable strong signal).

\subsubsection{Opportunity for laser transponders}

LLR is one of the most modern and exotic observational disciplines within astrometry, being used routinely for a host of fundamental astronomical and astrophysical studies.  However, even after more than 30 years of routine observational operation, LLR remains a non-trivial, sophisticated, highly technical, and remarkably challenging task.  Signal loss, proportional to the inverse 4th power of the Earth-Moon distance, but also the result of optical and electronic inefficiencies in equipment, array orientation, and heating, still requires that one observe mostly single photoelectron events.  Raw timing precision is some tens of picoseconds with the out-and-back range accuracy being approximately an order of magnitude larger.  Presently, we are down to sub-cm lunar ranging accuracies.  In this day of routine SLR operations, it is a sobering fact to realize that ranging to the Moon is many orders of magnitude harder than to an Earth-orbiting spacecraft.  Laser transponders may help to solve this problem.
Simple time-of-flight laser transponders offer a unique opportunity to overcome the problems above.  Although there are great opportunities for scientific advances provided by these instruments, there are also design challenges as transponders require power, precise pointing, and thermal stability.    

Active laser transponders on the lunar surface are attractive because of the strong return and insensitivity to lunar orientation effects.  A strong return would allow artificial satellite ranging stations to range the Moon.  
%Transponders may be used for ranging to more distant bodies than the Moon.  A lunar installation would provide valuable experience on their operational characteristics. 
%
%Active transponders would require power and would have more limited lifetimes than passive reflectors.  Transponders might have internal electronic delays that would need to be calibrated or estimated, and these delays might be temperature sensitive.  An unknown temperature variation would correlate with the EP test.  Transponders can also be used to good effect in asynchronous mode\cite{Degnan_2002}, wherein the received pulse train is not related to the transmitted pulse train, but the transponder unit records the temporal offsets between the two signals.
%
However, transponders require development: optical transponders detect a laser pulse and fire a return pulse back toward the Earth.\cite{Degnan_1993}  They give a much brighter return signal accessible to more stations on Earth. Active transponders would require power and would have more limited lifetimes than passive reflectors.  Transponders might have internal electronic delays that would need to be calibrated or estimated, since if these delays were temperature sensitive that would correlate with the SEP test.  Transponders can also be used to good effect in asynchronous mode,\cite{Degnan_2002,Degnan_2006} wherein the received pulse train is not related to the transmitted pulse train, but the transponder unit records the temporal offsets between the two signals.  The LLR experience can help determine the optimal location on the Moon for these devices. 

In addition to their strong return signals and insensitivity to lunar orientation effects, laser transponders are also attractive due to their potential to become increasingly important part of space exploration efforts.   Laser transponders on the Moon can be a prototype demonstration for later laser ranging to Mars and other celestial bodies to give strong science returns in the areas similar to those investigated with LLR.  A lunar installation would provide a valuable operational experience.

\section{Summary}
\label{sec:sum}

In this paper we considered the LLR tests of the equivalence principle (EP) performed with the Earth and Moon. If the ratio of gravitational mass to inertial mass is not constant, then there would be profound consequences for gravitation.  Such a violation of the EP would affect how bodies move under the influence of gravity.  The EP is not violated for Einstein's general theory of relativity, but violations are expected for many alternative theories of gravitation.  Consequently, tests of the EP are important to the search for a new theory of gravity.  

%Section~\ref{sec:history} gives a description of the Lunar Laser Ranging (LLR) technique along with a brief history. 

We considered the EP in its two forms (Sec.~\ref{sec:ep}); the weak equivalence principle (WEP) is sensitive to composition while the strong equivalence principle (SEP) considers possible sensitivity to the gravitational energy of a body.  The main sensitivity of the lunar orbit to the equivalence principle comes from the acceleration of the Earth and Moon by the Sun.  Any difference in those accelerations due to a failure of the equivalence principle causes an anomalous term in the lunar range with the 29.53 d synodic period.  The amplitude would be proportional to the difference in the gravitational to inertial mass ratios for Earth and Moon.  Thus, lunar laser ranging is sensitive to a failure of the equivalence principle due to either the WEP or the SEP.  In the case of the SEP, any violation of the equivalence principle can be related to a linear combination of the parametrized post-Newtonian parameters $\beta$ and $\gamma$.  

We also discussed the data and observational influences on its distribution (Sec.~\ref{sec:data}).  The evolution of the data from decimeter to centimeter quality fits is illustrated.  The LLR data set shows a variety of selection effects which influence the data distribution.  Important influences include phase of the Moon, season, distance, time of day, elevation in the sky, and declination.  For the LLR-enabled EP tests, selection with phase of the Moon is an important factor.  

An accurate model and analysis effort is needed to exploit the lunar laser range data to its full capability.  The model is the basis for the computer code that processes the range data (Sec.~\ref{sec:model}).  
%Section~\ref{sec:model} presents an overview of the model and computation.  
Further modeling efforts will be necessary to process range data of millimeter quality.  Two small effects for future modeling, thermal expansion and solar radiation pressure, are briefly discussed.  

Solutions for any EP violation are given in Section~\ref{sec:data_analysis}.  Several approaches to the solutions are used as checks.  The EP solution parameter can be either a ratio of gravitational to inertial masses or as a coefficient of a synodic term in the range equation.  The results are compatible in value and uncertainty.  Because $GM_{\tt Earth+Moon}$ correlates with the EP due to lunar phase selection effects, solutions are also made with this quantity fixed to a value based on non-LLR determinations of $GM_{Earth}$ and Earth/Moon mass ratio.  In all, five EP solutions are presented in Table~\ref{tab:1} and four are carried forward into Table~\ref{tab:2}.  As a final check, spectra of the post-fit residuals from a solution without any EP solution parameter are examined for evidence of any violation of the EP.  No such signature is evident.  The analysis of the LLR data does not show significant evidence for a violation of the EP compared to its uncertainty.  The final result for $[(M_G/M_I)_E -(M_G/M_I)_M]_{EP}$ is $(-1.0 \pm 1.4) \times 10^{-13}$.  

To gain insight into the lunar orbit and the solution for the EP, short trigonometric series expansions are given for the lunar orbit and orientation which are appropriate for a range expansion.  This is used to show how the data selection with lunar phase correlates the EP solution parameter with $GM_{\tt Earth+Moon}$.  To separate these and other relevant parameters, one wishes a good distribution of observations with lunar phase, orbital mean anomaly and argument of latitude, and, as seen from Earth, hour angle and declination.  

The result for the SEP is derived (subsection~\ref{sec:sep_solution}) from the total value determined by LLR by subtracting the laboratory result for the WEP determined at the University of Washington.  The Moon has a small core while the Earth has a large iron rich core.  Both have silicate mantles.  The WEP sensitivity of the Moon depends most strongly on the difference in iron content between the two bodies.  The SEP result is $[(M_G/M_I)_E -(M_G/M_I)_M]_{\tt SEP} = (-2.0 \pm 2.0) \times 10^{-13}$, which we do not consider to be a significant difference from the zero of general relativity.  

The SEP test can be related to the parametrized post-Newtonian (PPN) parameters $\beta$ and $\gamma$ (subsection~\ref{sec:beta}).  For conservative theories of relativity, one gets $4\beta - \gamma - 3 = (4.4\pm4.5)\times 10^{-4}$.  The Cassini spacecraft result for $\gamma$ allows a value for $\beta$ to be extracted.  That result is $\beta - 1 = (1.2 \pm 1.1) \times 10^{-4}$, which is the most accurate determination to date.  Again, we do not consider this $\beta$ value to be a significant deviation from the unity of general relativity.  

Finally, we discussed the efforts that are underway to extend the accuracies to millimeter levels (Sec.~\ref{sec:ememrging_oops}).  The expected improvement in the accuracy of LLR tests of gravitational physics expected with extended data set with existing stations and also with a new APOLLO instrument will bring significant new insights to our understanding of the fundamental physics laws that govern the evolution of our universe. The scientific results are very significant which justifies the nearly 40 years of history of LLR research and technology development.  

The lunar laser ranging results in this paper for the equivalence principle, strong equivalence principle, and PPN $\beta$ are consistent with the expectations of Einstein's general theory of relativity.  It is remarkable that general relativity has survived a century of testing and that the equivalence principle is intact after four centuries of scrutiny.  Each new significant improvement in accuracy is unknown territory and that is reason for future tests of the equivalence principle.  

\section*{Acknowledgments}
 
We acknowledge and thank the staffs of the Observatoire de la C\^ote d'Azur, Haleakala, and University of Texas McDonald ranging stations.  The analysis of the planetary data was performed by E. Myles Standish. The research described in this paper was carried out at the Jet Propulsion Laboratory, California Institute of Technology, under a contract with the National Aeronautics and Space Administration.

%===================================

%***************************


\begin{thebibliography}{145}

\bibitem%[AFCRL(1969)]
{AFCRL_1969}
Air Force Cambridge Research Laboratories, Bull. G\'eod\'esique 94, 443-444 (1969).

\bibitem%[Abalakin, Kokurin(1981)]
{Abalakin-Kokurin-1981} 
Abalakin, V. K., Kokurin, Yu. L.,
``Optical detection and ranging of the moon.'' Usp. Fiz. Nauk 134, 526-535 (1981). 

\bibitem%[Adelberger et al.(1990a)]
{Adelberger_etal_1990a} 
Adelberger, E. G., Heckel, B. R., Smith, G., Su, Y., and  Swanson, H. E.,``E\"otv\"os experiments, lunar ranging, and the strong equivalence principle,'' Nature 347, 261-263 (1990). 

\bibitem%[Adelberger et al.(1990b)]
{Adelberger_etal_1990b} 
Adelberger, E. G., Stubbs, C.~W., Heckel, B.~R., Smith, G., Su, Y., Swanson, H. E., Smith, G., Gundlach, J. H., and Rogers, W. F., ``Testing the equivalence principle in the field of the Earth: particle physics at masses below 1 µeV?'' Phys. Rev. D, 42, 3267-3292 (1990).

\bibitem%[Adelberger(2001)]
{Adelberger_2001}
Adelberger, E. G., ``New Tests of Einstein's Equivalence Principle and Newton's inverse-square law,'' Class. Quantum Grav. 18, 2397-2405 (2001).

\bibitem%[Alley(1972)]
{Alley_1972} 
Alley, C. O., ``Story of the development of the Apollo 11 laser ranging retro-reflector experiment,'' Adventures in Experimental Physics, ed. by B. Maglich, 132-149 (1972).

\bibitem%[Anderson et al.(1996)]
{Anderson_etal_1996} 
Anderson, J. D., Gross, M., Nordtvedt, K. L., and Turyshev, S. G., ``The Solar Test of the Equivalence Principle,'' Astrophys. Jour. 459, 365-370 (1996).

\bibitem%[Anderson and Williams(2001)]
{Anderson_Williams_2001} 
Anderson, J. D., and Williams, J. G., ``Long-Range Tests of the Equivalence Principle,'' Class. Quantum Grav. 18, 2447-2456 (2001).

\bibitem%[Bae{\ss}ler et al.(1999)]
{Baessler_etal_1999} 
Bae{\ss}ler, S., Heckel, B., Adelberger, E. G., Gundlach, J., Schmidt, U., and Swanson, E., ``Improved Test of the Equivalence Principle for Gravitational Self-Energy,'' Phys. Rev. Lett. 83, 3585-3588 (1999).

\bibitem%[Bertotti, Iess, and Tortora(2003)]
{Bertotti_etal_2003}
Bertotti, B., Iess, L., and Tortora, P., ``A test of general relativity using radio links with the Cassini spacecraft,'' Nature 425, 374-376 (2003).

\bibitem%[Bender et al.(1973)]
{Bender_etal_1973}
Bender, P. L., Currie, D. C., Dicke, R. H., Eckhardt, D. H., Faller, J. E., Kaula, W. M., Mulholand, J. D., Plotkin, H. H., Poultney, S. K., Silverberg, E. C., Wilkinson, D. T., Williams, J. G., and Alley, C. O., ``The Lunar Laser Ranging Experiment,'' Science 182, 229-237 (1973).

\bibitem%[Bod et al.(1991)]
{Bod_etal_1991}
Bod, L., Fischbach, E. Marx, G. and N\'aray-Ziegler, M., ``One Hundred Years of the E\"otv\"os Experiment,'' Acta Physica Hungarica 69, 335-355 (1991). 

\bibitem%[Braginsky and Panov(1972)]
{Braginsky_Panov_1972}
Braginsky, V. B., and Panov, V. I., ``Verification of Equivalence Principle of Inertial and Gravitational Mass,'' Zh. Eksp. Teor. Fiz. 61, 873-876 (1971), [Sov. Phys. JETP 34, 463-466 (1972)].

\bibitem%[Braginsky, Gurevich, and Zybin(1992)]
{Braginsky_etal_1992}
Braginsky, V. B., Gurevich, A. V., and Zybin, K. P., ``The influence of dark matter on the motion of planets and satellites in the solar system,'' Phys. Lett. A 171, 275-277 (1992).

\bibitem%[Braginsky(1994)]
{Braginsky_1994}
Braginsky, V. B., ``Experimental gravitation (what is possible and what is interesting to measure).'' Class. Quantum Grav. 11, A1-A7 (1994).

\bibitem%[Calame et al.(1970)]
{Calame_etal_1970} 
Calame, O., Fillol, M.-J., Guérault, G., Muller, R., Orszag, A., Pourny, J.-C., R\"osch, J., and de Valence, Y., ``Premiers \'echos lumineux sur la lune obtenus par le t\'el\'em\`etre du Pic du Midi,'' Comptes Rendus Acad. Sci. Paris, Ser. B 270, 1637-1640 (1970).

\bibitem%[Chandler, Reasenberg, and Shapiro(1994)]
{Chandler_etal_1994} 
Chandler, J. F., Reasenberg, R. D., and Shapiro, I. I., ``New results on the Principle of Equivalence,'' Bull. Am. Astron. Soc. 26, 1019 (1994).

\bibitem%[Chapront-Touz\'e and Chapront(1988)]
{Chapront-Touze_Chapront_1988} 
Chapront-Touz\'e, M., and Chapront, J., ``ELP 2000-85: a semi-analytical lunar ephemeris adequate for historical times,'' Astron. Astrophys. 190, 342-352 (1988).  

\bibitem%[Chapront-Touz\'e and Chapront(1991)]
{Chapront-Touze_Chapront_1991}
Chapront-Touz\'e, M., and Chapront, J., Lunar Tables and Programs from 4000 B. C. to A. D. 8000 (Willmann-Bell, Richmond, 1991).

\bibitem%[Chapront, Chapront-Touz\'e, and Francou(2002)]
{Chapront_etal_2002} 
Chapront, J., Chapront-Touz\'e, M., and Francou, G., ``A new determination of lunar orbital parameters, precession constant and tidal acceleration from LLR measurements,'' Astron. Astrophys. 387, 700-709 (2002).  

\bibitem%[Damour(1996)]
{Damour_1996}  
Damour, T., ``Testing the Equivalence Principle: why and how?'' Class. Quantum Grav. 13, A33-A42 (1996).

\bibitem%[Damour(2001)]
{Damour_2001}  
Damour, T., ``Questioning the Equivalence Principle,'' (2001) [arXiv:gr-qc/0109063].

\bibitem%[Damour and Esposito-Far\`ese(1996a)]
{Damour_Esposito-Farese_1996a}
Damour, T., and Esposito-Far\`ese, G., ``Testing gravity to second post-Newtonian order: a field-theory approach,'' Phys. Rev. D 53, 5541-5578 (1996a).

\bibitem%[Damour and Esposito-Far\`ese(1996b)]
{Damour_Esposito-Farese_1996b}
Damour, T., and Esposito-Far\`ese, G., ``Tensor-scalar gravity and binary-pulsar experiments,'' Phys. Rev. D, 54, 1474-1491 (1996b).

\bibitem%[Damour and Nordtvedt(1993a)]
{Damour_Nordtvedt_1993a}
Damour, T., and Nordtvedt, K., Jr., ``General Relativity as a Cosmological Attractor of Tensor Scalar Theories,'' 
Phys. Rev. Lett. 70, 2217-2219 (1993a).

\bibitem%[Damour and Nordtvedt(1993b)]
{Damour_Nordtvedt_1993b}
Damour, T., and Nordtvedt, K., Jr., ``Tensor-scalar cosmological models and their relaxation toward general relativity,'' Phys. Rev. D, 48, 3436-3450 (1993b).

\bibitem%[Damour and Polyakov(1994a)]
{Damour_Polyakov_1994a}
Damour, T., and Polyakov, A. M., ``String Theory and Gravity,'' General Relativity Gravit. 26, 1171-1176 (1994a).

\bibitem%[Damour and Polyakov(1994b)]
{Damour_Polyakov_1994b}
Damour, T., and Polyakov, A. M., ``The string dilaton and a least coupling principle,'' Nucl. Phys. B423, 532-558 (1994b).

\bibitem%[Damour, Piazza, and Veneziano(2002a)]
{Damour_etal_2002a}
Damour, T., Piazza, F., and Veneziano, G., ``Runaway dilaton and equivalence principle violations,'' Phys. Rev. Lett. 89, 081601 (2002a) [arXiv:gr-qc/0204094].

\bibitem%[Damour, Piazza, and Veneziano(2002b)]
{Damour_etal_2002b}
Damour, T., Piazza, F., and Veneziano, G., ``Violations of the equivalence principle in a dilaton-runaway scenario,''
Phys. Rev. D 66, 046007 (2002b) [arXiv:hep-th/0205111].

\bibitem%[Damour and Sch\"afer(1991)]
{Damour_Schafer_1991}
Damour, T., and Sch\"afer, G., ``New tests of the strong equivalence principle using Binary-Pulsar data,''
Phys. Rev. Lett. 66, 2549-2552 (1991).

\bibitem%[Damour and Vokrouhlicky(1996a)]
{Damour_Vokrouhlicky_1996a}
Damour, T., and Vokrouhlicky, D., ``Equivalence Principle and the Moon,'' Phys. Rev. D 53, 4177-4201 (1996a).

\bibitem%[Damour and Vokrouhlicky(1996b)]
{Damour_Vokrouhlicky_1996b}
Damour, T., and Vokrouhlicky, D., ``Testing for gravitationally preferred directions using the lunar orbit,'' Phys. Rev. D 53, 6740-6740 (1996b).

\bibitem%[Degnan(1985)]
{Degnan_1985}
Degnan, J. J., ``Satellite Laser Ranging: Status and Future Prospects,'' IEEE Trans. Geosci. and Rem. Sens., Vol. GE-23, 398-413 (1985).

\bibitem%[Degnan(1993)]
{Degnan_1993}
Degnan, J. J., ``Millimeter accuracy satellite laser ranging: a review,'' Contributions of Space Geodesy to Geodynamics: Technology, Geodynamics Series, D.E. Smith and D.L. Turcotte (Eds.), AGU Geodynamics Series 25, 133-162 (1993).

\bibitem%[Degnan(2002)]
{Degnan_2002}
Degnan, J. J., ``Asynchronous Laser Transponders for Precise Interplanetary Ranging and Time Transfer,'' Journal of Geodynamics (Special Issue on Laser Altimetry), 551-594, (2002). 

\bibitem%[Degnan(2006)]
{Degnan_2006}  
J.\,J. Degnan, ``Laser Transponders for High Accuracy Interplanetary Laser Ranging and Time Transfer''. In {\it Lasers, Clocks, and Drag-Free: Exploration of Relativistic Gravity in Space}, eds. H. Dittus, C. Lammerzahl, and S.\,G. Turyshev,  pp. 231-242, (Springer, New York, 2006).

\bibitem%[Dickey, Newhall, and Williams(1989)]
{Dickey_Newhall_Williams_1989}
Dickey, J. O., Newhall, X X, and Williams, J. G., ``Investigating Relativity Using Lunar Laser Ranging: Geodetic Precession and the Nordtvedt Effect,'' Adv. Space Res. 9(9), 75-78 (1989).

\bibitem%[Dickey et al.(1994)]
{Dickey_etal_1994}
Dickey, J. O., Bender, P. L., Faller, J. E., Newhall, X X, Ricklefs, R. K., Shelus, P. J., Veillet, C., Whipple, A. L., Wiant, J. R., Williams, J. G., and Yoder, C. F., ``Lunar Laser Ranging: A Continuing Legacy of the Apollo Program,'' Science 265, 482-490 (1994).

\bibitem%[Eckhardt(1990)]
{Eckhardt_1990} 
Eckhardt, D. H., ``Gravitational shielding,'' Phys. Rev. D 42, 2144-2145 (1990).

\bibitem%[E\"otv\"os(1890)]
{Eotvos_1890} 
E\"otv\"os, R. v., Mathematische und Naturwissenschaftliche Berichte aus Ungarn 8, 65 (1890).

\bibitem%[E\"otv\"os, Pek\'ar, and Fekete(1922)]
{Eotvos_etal_1922}
E\"otv\"os, R. v.,  Pek\'ar, D., Fekete, E., Annalen der Physik (Leipzig) 68, 11, 1922. English translation for the U. S. Department of Energy by J. Achzenter, M. Bickeb\"oller, K. Br\"auer, P. Buck, E. Fischbach, G. Lubeck, C. Talmadge, University of Washington preprint 40048-13-N6. - More complete English text reprinted earlier in Annales Universitatis Scientiarium Budapestiensis de Rolando E\"otv\"os Nominate, Sectio Geologica 7, 111 (1963).

\bibitem%[Faller et al.(1969)]
{Faller_etal_1969} 
Faller, J. E., Winer, I., Carrion, W., Johnson, T. S.,  Spadin, P., Robinson, L., Wampler, E. J., and Wieber, D., ``Laser beam directed at the lunar retro-reflector array: observations of the first returns,'' Science 166, 99-102 (1969).

\bibitem%[Ferrari et al.(1980)]
{Ferrari_etal_1980} 
Ferrari, A. J., Sinclair, W. S., Sjogren, W. L., Williams, J. G. and Yoder, C. F., ``Geophysical Parameters of the Earth-Moon System,'' J. Geophys. Res. 85, 3939-3951 (1980).

\bibitem%[Flasar and Birch(1973)]
{Flasar_Birch_1973} 
Flasar, F. M., and Birch, F., ``Energetics of core formation: a correction,'' J. Geophys. Res. 78, 6101-6103 (1973).  

\bibitem%[Hood et al.(1999)]
{Hood_etal_1999}
Hood, L. L., Mitchell, D. L., Lin, R. P., Acuna, M. H., and Binder, A. B., ``Initial measurements of the lunar induced magnetic dipole moment using Lunar Prospector magnetometer data,'' Geophys. Res. Lett., 26, 2327-2330 (1999).

\bibitem%[Kokurin(2003)]
{Kokurin_2003}
Kokurin, Yu. I, ``Lunar laser ranging: 40 years of research,'' Quantum Electronics 33(1), 45-47 (2003).

\bibitem%[Konopliv et al.(1998)]
{Konopliv_etal_1998}
Konopliv, A. S., Binder, A. B., Hood, L. L., Kucinskas, A. B., Sjogren, W. L., and Williams, J. G., ``Improved gravity field of the Moon from Lunar Prospector,'' Science 281, 1476-1480 (1998).

\bibitem%[Konopliv et al.(2002)]
{Konopliv_etal_2002}
Konopliv, A. S., Miller, J. K., Owen, W. M., Yeomans, D. K., and  Giorgini, J. D., ``A Global Solution for the Gravity Field, Rotation, Landmarks, and Ephemeris of Eros,'' Icarus 160, 289–299 (2002).

\bibitem%[Kozai(1972)]
{Kozai_1972}
Kozai, Y., ``Lunar laser ranging experiments in Japan,'' Space Research XII, 211-217, (1972).

\bibitem%[Kuskov and Kronrod(1998a)]
{Kuskov_Kronrod_1998a}
Kuskov, O. L., and Kronrod, V. A., ``A model of the chemical differentiation of the Moon,'' Petrology 6, 564-582 (1998a).  

\bibitem%[Kuskov and Kronrod(1998b)]
{Kuskov_Kronrod_1998b}
Kuskov, O. L., and Kronrod, V. A., ``Constitution of the Moon, 5, Constraints on composition, density, temperature, and radius of a core,'' Phys. Earth Planet. Inter., 107, 285-306 (1998b).  

\bibitem%[Larimer(1986)]
{Larimer_1986}
Larimer, J. W., ``Nebular chemistry and theories of lunar origin, in Origin of the Moon,'' edited by W. K. Hartmann, R. J. Phillips, and G. J. Taylor, 145-171 (Lunar and Planet. Inst., Houston, Tex., 1986). 

\bibitem%[Lorimer and Freire(2004)]
{Lorimer_Freire_2004}
Lorimer, D. R., and Freire, P. C. C., ``New limits on the strong equivalence principle from two long-period circular-orbit binary pulsars,'' (2004) [arXiv:astro-ph/0404270].

\bibitem%[Luck, Miller, and Morgan(1973)]
{Luck_etal_1973}
Luck, J., Miller, M. J., and Morgan, P. J., ``The National Mapping Lunar Laser Program,'' in proceedings of The Earth's Gravitational Field and Secular Variations in Position, a conference held 26-30 November, 1973 at New South Wales, Sydney, Australia. Australian Academy of Science and the International Association of Geodesy, 413 (1973).

\bibitem%[Majorana(1920)]
{Majorana_1920}
Majorana, Q., ``On gravitation. Theoretical and experimental researches,'' Philos. Mag. 39, 488-504 (1920).

\bibitem%[Marini and Murray(1973)]
{Marini_Murray_1973} 
Marini, J. W., Murray, C. W., Jr., ``Correction of Laser Range Tracking Data for Atmospheric Refraction at Elevation Angles Above 10 Degrees,'' NASA Technical Report, X-591-73-351 (1973).

\bibitem%[McCarthy and Petit(2003)]
{McCarthy_Petit_2003} 
McCarthy, D. D., and Petit, G. eds. ``IERS Conventions (2003)'' (2003). IERS Technical Note~\#32. Frankfurt am Main: Verlag des Bundesamts f\"ur Kartographie und Geod\"asie, 2004. 127 pp. Electronic version available at {\tt http://www.iers.org/iers/products/conv/}

\bibitem%[Morgan and King(1982)]
{Morgan_King_1982}  
Morgan, P., King, R. W., ``Determination of coordinates for the Orroral Lunar Ranging Station, in High-precision earth rotation and earth-moon dynamics: Lunar distances and related observations'' Proceedings of the Sixty-third Colloquium, Grasse, Alpes-Maritimes, France, May 22-27, 1981. (A82-47176 24-89) Dordrecht, D. Reidel Publishing Co., 305-311 (1982).

\bibitem%[M\"uller et al.(1996)]
{Mueller_etal_1996b}
M\"uller, J., Schnider, M., Soffel, M., and Ruder, H., ``Determination of Relativistic Quantities by Analyzing Lunar Laser Ranging Data,'' In proceedings of ``the Seventh Marcel Grossmann Meeting,'' World Scientific Publ., eds. R. T. Jantzen, G. M. Keiser, and R. Ruffini, 1517 (1996).

\bibitem%[M\"uller and Nordtvedt(1998)]
{Mueller_Nordtvedt_1998}
M\"uller, J., and Nordtvedt, K., Jr., ``Lunar laser ranging and the equivalence principle signal,'' Phys. Rev. D 58, 62001/1-13 (1998). 

\bibitem%[Murphy et al.(2000)]
{Murphy_etal_2000}
Murphy, T. M., Jr., Strasburg, J. D., Stubbs, C. W., Adelberger, E. G., Angle, J., Nordtvedt, K., Williams, J. G., Dickey, J. O., and Gillespie, B., ``The Apache Point Observatory Lunar Laser-Ranging Operation (APOLLO),'' Proceedings of 12th International Workshop on Laser, Ranging, Matera, Italy (November 2000)\\ http://www.astro.washington.edu/tmurphy/apollo/matera.pdf

\bibitem%[Murphy et al.(2007)]
{Murphy_etal_2007}
T. W. Murphy, Jr., E. L. Michelson, A. E. Orin, E. G. Adelberger, C. D. Hoyle, H. E. Swanson, C. W. Stubbs, J. E. Battat, 
``APOLLO: Next-Generation Lunar Laser Ranging'', Int. J. Mod. Phys. D 16(12a), 2127 (2007).

\bibitem%[Murphy et al.(2008)]
{Murphy_etal_2008}
T. W. Murphy, Jr., E. G. Adelberger, J.B.R. Battat, L.N. Carey, C.D. Hoyle, P. LeBlanc, E.L. Michelsen, K. Nordtvedt, A.E. Orin, J.D. Strasburg, C.W. Stubbs, H.E. Swanson, E. Williams, 
``APOLLO: the Apache Point Observatory Lunar Laser-ranging Operation: Instrument Description and First Detections'', 
Publ. Astron. Soc. Pac. 120(863), 20-37 (2008).

\bibitem%[Nordtvedt(1968a)]
{Nordtvedt_1968a}
Nordtvedt, K., Jr., ``Equivalence Principle for Massive Bodies.  I. Phenomenology,'' Phys. Rev. 169, 1014-1016 (1968a).

\bibitem%[Nordtvedt(1968b)]
{Nordtvedt_1968b}
Nordtvedt, K., Jr., ``Equivalence Principle for Massive Bodies. II. Theory,'' Phys. Rev. 169, 1017-1025 (1968b).

\bibitem%[Nordtvedt(1968c)]
{Nordtvedt_1968c}
Nordtvedt, K., Jr., ``Testing Relativity with Laser Ranging to the Moon,'' Phys. Rev. 170, 1186-1187 (1968c).

\bibitem%[Nordtvedt(1970)]
{Nordtvedt_1970}
Nordtvedt, K., Jr., ``Solar system E\"otv\"os experiments,'' Icarus 12, 91-100, (1970). 

\bibitem%[Nordtvedt(1991)]
{Nordtvedt_1991} 
Nordtvedt, K., Jr., ``Lunar Laser Ranging Re-examined: The Non-Null Relativistic Contribution,'' Phys. Rev. D 43, 3131-3135 (1991).

\bibitem%[Nordtvedt(1994)]
{Nordtvedt_1994}
Nordtvedt, K., Jr., ``Cosmic Acceleration of the Earth and Moon by Dark-Matter,'' Astroph. J. 437, 529-531 (1994).  

\bibitem%[Nordtvedt(1995)]
{Nordtvedt_1995} 
Nordtvedt, K., Jr., ``The relativistic orbit observables in lunar laser ranging,'' Icarus 114, 51-62 (1995).

\bibitem%[Nordtvedt(1998)]
{Nordtvedt_1998} 
Nordtvedt, K., Jr., ``Optimizing the observation schedule for tests of gravity in lunar laser ranging and similar experiments,'' Class. Quantum Grav. 15, 3363-3381 (1998).  

\bibitem%[Nordtvedt(1999)]
{Nordtvedt_1999} 
Nordtvedt, K., Jr., ``30 years of lunar laser ranging and the gravitational interaction,'' Class. Quantum Grav. 16, A101-A112 (1999).  

\bibitem%[Nordtvedt(2003)]
{Nordtvedt_2003} 
Nordtvedt, K., Jr., ``Lunar Laser Ranging - A Comprehensive Probe of Post-Newtonian Gravity,'' (2003) [arXiv:gr-qc/0301024].

\bibitem%[Nordtvedt, M\"uller, and Soffel(1995)]
{Nordtvedt_etal_1995}
Nordtvedt, K. L., M\"uller, J., and Soffel, M., ``Cosmic Acceleration of the Earth and Moon by Dark-Matter,'' Astron. Astrophysics 293, L73-L74 (1995). 

\bibitem%[Nordtvedt and Vokrouhlicky(1997)]
{Nordtvedt_Vokrouhlicky_1997}
Nordtvedt K., Jr., and Vokrouhlicky, D., ``Recent Progress in Analytical Modeling of the Relativistic Effects in the Lunar Motion,'' in `Dynamics and Astronomy of the Natural and Artificial Celestial Bodies', eds: I.M. Wytrzysczcak, J.~H. Lieske and R.~A. Feldman (Kluwer Academic Publishers, Dordrecht), 205 (1997).

\bibitem%[Orellana and Vucetich(1988)]
{Orellana_Vucetich_1988}
Orellana, R. B., and Vucetich, H., ``The principle of equivalence and the Trojan asteroids,'' Astron. Astrophys. 200, 248-254 (1988).  

\bibitem%[Orellana and Vucetich(1993)]
{Orellana_Vucetich_1993}
Orellana, R. B., and Vucetich, H., ``The Nordtvedt Effect in the Trojan Asteroids,'' Astron. Astrophys 273, 313-317 (1993).

\bibitem%[Roll, Krotkov, and Dicke(1964)]
{Roll_etal_1964}
Roll, P. G., Krotkov, R., and Dicke, R. H., ``The Equivalence Principle of Inertial and Gravitational Mass,'' Ann. Phys. (N.Y.) 26, 442-517 (1964).

\bibitem%[Ries et al.(1992)]
{Ries_etal_1992}
Ries, J. C., Eanes, R. J., Shum, C. K., and Watkins, M. M., ``Progress in the determination of the gravitational coefficient of the Earth,''  Geophys. Res. Lett. 19, 529-531 (1992).

\bibitem%[Henry Norris Russell(1921)]
{Russell_1921} 
Russell, H. N., ``On Majorana's theory of gravitation,'' Astrophys. J. 54, 334-346 (1921).

\bibitem%[Samain et al.(1998)]
{Samain_etal_1998} 
Samain, E., Mangin, J. F., Veillet, C., Torre, J. M., Fridelance, P., Chabaudie, J. E., Feraudy, D., Glentzlin, M., Pham Van, J., Furia, M., Journet, A., and Vigouroux, G., ``Millimetric Lunar Laser Ranging at OCA (Observatoire de la C\^ote d'Azur),'' Astron. Astrophys. Suppl. Ser. 130, 235-244 (1998).

\bibitem%[Singe(1960)]
{Singe_1960}  
Singe, J. L., Relativity: the General Theory (Amsterdam: North-Holland, 1960).

\bibitem%[Smith et al.(1993)]
{Smith_etal_1993} 
Smith, G., Adelberger, E. G., Heckel, B. R., Su, Y., ``Test of the equivalence principle for ordinary matter falling toward dark matter,'' Phys. Rev. Lett. 70, 123-126 (1993).

\bibitem%[Shapiro, Counselman, and King(1976)]
{Shapiro_etal_1976} 
Shapiro, I. I., Counselman, C. C., III, and King, R. W., ``Verification of the Principle of Equivalence for Massive Bodies,'' Phys. Rev. Lett. 36, 555-558 (1976).

\bibitem%[Shelus et al.(2003)]
{Shelus_etal_2003}
Shelus, P., Ries, J. G., Wiant, J. R., Ricklefs, R. L., ``McDonald Ranging: 30 Years and Still Going,'' in Proc. of 13th International Workshop on Laser Ranging, October 7-11, 2002, Washington, D. C. (2003), http://cddisa.gsfc.nasa.gov/lw13/lw$\underline{ }$proceedings.html

\bibitem%[Standish and Williams(2005)]
{Standish_Williams_2005}
Standish, E. M., and Williams, J. G., ``Orbital Ephemerides of the Sun, Moon, and Planets,'' Chapter 8 of the Explanatory Supplement to the American Ephemeris and Nautical Almanac, in press (2005). 

\bibitem%[Standish(1998)]
{Standish_1998}
Standish, E. M., ``Time scales in the JPL and CfA ephemerides,'' Astron. Astrophys. 336, 381-384 (1998).

\bibitem%[Su et al.(1994)]
{Su_etal_1994} 
Su, Y., Heckel, B. R., Adelberger, E. G., Gundlach, J. H., Harris, M., Smith, G. L., and Swanson, H. E., ``New tests of the universality of free fall,'' Phys. Rev. D 50, 3614-3636 (1994).

\bibitem%[Tremaine(1992)]
{Tremaine_1992}
Tremaine, S., ``The Dynamical Evidence for Dark Matter,'' Physics Today 45, 28-36 (1992).

\bibitem%[Turyshev et al.(2004)]
{Turyshev_etal_2004} 
Turyshev, S. G., Williams, J. G., Nordtvedt, K., Jr., Shao, M., Murphy, T. W., Jr., ``35 Years of Testing Relativistic Gravity: Where do we go from here?'', in Proc. ``302.WE-Heraeus-Seminar: Astrophysics, Clocks and Fundamental Constants, 16-18 June 2003. The Physikzentrum, Bad Honnef, Germany.'' Springer Verlag, Lect. Notes Phys.  648, 301-320, (2004) [arXiv:gr-qc/0311039].

\bibitem{Turyshev_etal_2007} 
S. G. Turyshev, U. E. Israelsson, M. Shao, N. Yu, A. Kusenko, E. L. Wright, C.W.F. Everitt, M. Kasevich, J. A. Lipa, J. C. Mester, R. D. Reasenberg, R. L. Walsworth, N. Ashby, H. Gould, H. J. Paik, 
``Space-based research in fundamental physics and quantum technologies,'' {\it Inter. J. Modern Phys. D \bf 16}(12a), 1879-1925 (2007), arXiv:0711.0150 [gr-qc]

\bibitem{Turyshev-Williams-2007} 
S. G. Turyshev and J. G. Williams, 
``Space-based tests of gravity with laser ranging,'' {\it Int. J. Mod. Phys. D \bf 16}(12a), 2165-2179 (2007) [arXiv:gr-qc/0611095]

\bibitem{Turyshev_2008} 
Turyshev, S. G., 
``Experimental Tests of General Relativity,'' {\it Annu. Rev. Nucl. Part. Sci. \bf 58}, 207-248 (2008), arXiv:0806.1731 [gr-qc]. 

\bibitem%[Ulrich(1982)]
{Ulrich_1982}
Ulrich, R. K., ``The Influence of Partial Ionization and Scattering States on the Solar Interior Structure,'' Astrophys. J. 258, 404-413 (1982).

\bibitem%[Veillet et al.(1993)]
{Veillet_etal_1993} 
Veillet, C., J. F. Mangin, J. E. Chabaudie, C. Dumoulin, D. Feraudy, and J. M. Torre, ``Lunar laser ranging at CERGA for the ruby period (1981-1986),'' in Contributions of Space Geodesy to Geodynamics: Technology, AGU Geodynamics Series, 25, edited by D. E. Smith and D. L. Turcotte, 133-162 (1993).

\bibitem%[Vokrouhlicky(1997)]
{Vokrouhlicky_1997}
Vokrouhlicky, D., ``A note on the solar radiation perturbations of lunar motion,'' Icarus 126, 293-300 (1997).

\bibitem%[Wex(2001)]
{Wex_2001}
Wex, N., ``Pulsar timing - strong gravity clock experiments,'' in Gyros, Clocks, and Interferometers: Testing Relativistic Gravity in Space.  C. L\"ammerzahl et al., eds.,  Lecture Notes in Physics 562, 381-399 (Springer, Berlin 2001).

\bibitem%[Will(1971)]
{Will_1971}
Will, C. M., ``Theoretical Frameworks for Testing Relativistic Gravity. II. Parametrized Post-Newtonian Hydrodynamics, and the Nordtvedt Effect,'' Astrophys. J., 163, 611-628 (1971).

\bibitem%[Will and Nordtvedt(1972)]
{Will_Nordtvedt_1972}
Will, C. M. and Nordtvedt, K., Jr., ``Conservation Laws and Preferred Frames in Relativistic Gravity 1: Preferred-Frame Theories and an Extended PPN Formalism,'' Astrophys. J. 177, 757-774 (1972).

\bibitem%[Will(1990)]
{Will_1990}
Will, C. M., ``General Relativity at 75: How Right was Einstein?'', Science 250, 770-771 (1990).

\bibitem%[Will(1993)]
{Will_1993} 
Will, C. M., Theory and Experiment in Gravitational Physics (Cambridge, 1993).

\bibitem%[Will(2001)]
{Will_2001}
Will, C. M., ``The Confrontation between General Relativity and Experiment,'' Living Rev. Rel. 4, 4 (2001) [arXiv:gr-qc/0103036].

\bibitem%[Williams et al.(1976)]
{Williams_etal_1976} 
Williams, J. G., Dicke, R. H., Bender, P. L., Alley, C. O., Carter, W. E., Currie, D. G., Eckhardt, D. H., Faller, J. E., Kaula, W. M., Mulholland, J. D., Plotkin, H. H., Poultney, S. K., Shelus, P. J., Silverberg, E. C., Sinclair, W., S., Slade, M. A., and Wilkinson, D. T., ``New Test of the Equivalence Principle from Lunar Laser Ranging,'' Phys. Rev. Lett. 36, 551-554 (1976).

\bibitem%[Williams, Newhall, and Dickey(1996a)]
{Williams_etal_1996a}
Williams, J. G., Newhall, X X, and Dickey, J. O., ``Relativity Parameters Determined from Lunar Laser Ranging,'' Phys. Rev. D 53, 6730-6739 (1996).

\bibitem%[Williams, Newhall, and Dickey(1996b)]
{Williams_etal_1996b}
Williams, J. G., Newhall, X X, and Dickey, J. O., ``Relativity parameters determined from lunar laser ranging,'' In the Proc. of ``The Seventh Marcel Grossmann meeting on recent developments in theoretical and experimental general relativity, gravitation, and relativistic field theories,'' Stanford University, 24-30 July 1994, ed. R. T. Jantzen and G. M. Keiser, World Scientific, Singapore, 1529-1530 (1996).

\bibitem%Williams et al.(2001b)]
{Williams_etal_2001b}
Williams, J. G., Boggs, D. H., Yoder, C. F., Ratcliff, J. T., and Dickey, J. O., ``Lunar rotational dissipation in solid body and molten core,'' J. Geophys. Res. Planets 106, 27933-27968 (2001).

\bibitem%[Williams et al.(2002)]
{Williams_etal_2002}
Williams, J. G., Boggs, D. H.,  Dickey, J. O., and Folkner, W. M., ``Lunar Laser Tests of Gravitational Physics,'' in proceedings of The Ninth Marcel Grossmann Meeting, World Scientific Publ., eds. V. G. Gurzadyan, R. T. Jantzen, and R. Ruffini, 1797-1798 (2002).

\bibitem%[Williams and Dickey(2003)]
{Williams_Dickey_2003}
Williams, J. G. and Dickey, J. O., ``Lunar Geophysics, Geodesy, and Dynamics,'' in proc. of 13th International Workshop on Laser Ranging, October 7-11, 2002, Washington, D. C. (2003),  {\tt http://cddisa.gsfc.nasa.gov/lw13/lw$\underline{ }$proceedings.html}

\bibitem%[Williams, Turyshev, and Murphy(2004)]
{Williams_Turyshev_Murphy_2004} 
Williams, J. G., Turyshev, S. G., Murphy, T. W., Jr., ``Improving LLR Tests of Gravitational Theory,'' International Journal of Modern Physics D 13, 567-582 (2004) [arXiv:gr-qc/0311021].

\bibitem%[Williams, Turyshev, and Boggs(2004)]
{Williams_Turyshev_Boggs_2004} 
Williams, J. G., Turyshev, S. G., Boggs, D. H., ``Progress in Lunar Laser Ranging Tests of Relativistic Gravity,'' Phys. Review Letters 93, 261101 (2004) [arXiv:gr-qc/0411113]. 

\bibitem%[Williams et al.(2005)]
{Williams_etal_2005} 
Williams, J. G., Turyshev, S. G., Boggs, D. H., and Ratcliff, J.T., ``Lunar Laser Ranging Science: Gravitational Physics and Lunar Interior and Geodesy,'' %35th COSPAR Scientific Assembly, July 18-24, 2004, Paris, France, in 
{Adv. Space Res. \bf 37}(1), 67-71 (2006), arXiv:gr-qc/0412049.

%James G. Williams, Slava G. Turyshev, Dale H. Boggs, and J. Todd Ratcliff, Lunar Laser Ranging Science: Gravitational Physics and Lunar Interior and Geodesy, 35th COSPAR Scientific Assembly, July 18-24, 2004, Paris, France, in Advances in Space Research, vol. 37, Issue 1, The Moon and Near-Earth Objects, 67-71, doi: 10.1016/j.asr. 2005.05.13, (2006). [arXiv:gr-qc/0412049]

\bibitem%[Williams(2005)]
{Williams_2005} 
Williams, J. G., ``Solar System Tides - Formulation and Application to the Moon,'' in preparation (2009). 

\end{thebibliography}
\end{document}